\documentclass[twocolumn,aps,prb,superscriptaddress]{revtex4-2}
\usepackage[T1]{fontenc}
\usepackage{babel}
\usepackage{amsmath}
\usepackage{amsfonts}
\usepackage{graphicx}
\usepackage{braket}
\usepackage{multirow}
\usepackage[unicode=true]
 {hyperref}

\makeatletter
\renewcommand{\fnum@figure}{FIG.~\thefigure}
\renewcommand{\theequation}{\arabic{equation}}
\usepackage{algorithm,algpseudocode}
\usepackage{color,xcolor}
\usepackage[caption=false]{subfig}
\definecolor{darkRed}{RGB}{130,0,0}
\definecolor{darkGreen}{RGB}{0,130,0}
\definecolor{darkBlue}{RGB}{0,0,130}
\usepackage{hyperref}

\hypersetup{
      colorlinks=true,
      citecolor=darkGreen,
      linkcolor=darkGreen,
      urlcolor=darkGreen}

\makeatother

\begin{document}

\title{Stability of quantum chaos against weak non-unitarity}

\author{Yi-Cheng Wang}\affiliation{Department of Physics, University of California, Berkeley, CA 94720, USA}
\author{Ehud Altman}\affiliation{Department of Physics, University of California, Berkeley, CA 94720, USA}
\affiliation{Materials Sciences Division, Lawrence Berkeley National Laboratory, Berkeley, CA 94720, USA}
\author{Samuel J. Garratt}
\affiliation{Department of Physics, University of California, Berkeley, CA 94720, USA}
\affiliation{Department of Physics, Princeton University, NJ 08544, USA}

\date{\today}

\begin{abstract}
We study the quantum dynamics generated by the repeated action of a non-unitary evolution operator on a system of qubits. Breaking unitarity can lead to the purification of mixed initial states, which corresponds to the loss of sensitivity to initial conditions, and hence the absence of a key signature of dynamical chaos. However, the scrambling of quantum information can delay purification to times that are exponential in system size. Here we study purification in systems whose evolution operators are fixed in time, where all aspects of the dynamics are in principle encoded in spectral properties of the evolution operator for a single time step. The operators that we study consist of global Haar random unitary operators and non-unitary single-qubit operations. We show that exponentially slow purification arises from a distribution of eigenvalues in the complex plane that forms a ring with sharp edges at large radii, with the eigenvalue density exponentially large near these edges. We argue that the sharp edges of the eigenvalue distribution arise from level attraction along the radial direction in the complex plane. By calculating the spectral form factor we also show that there is level repulsion around the azimuthal direction, even close to the outer edge of the ring of eigenvalues. Our results connect this spectral signature of quantum chaos to the sensitivity of the system to its initial conditions.
\end{abstract}

\maketitle

\section{Introduction}

Chaos is the extreme sensitivity of the time-evolved state of a system to details of the initial state \cite{Gutzwiller1990}. In generic many-body quantum systems this sensitivity arises from the unitarity of time evolution, which allows the state to explore a Hilbert space whose dimension is exponential in system size while preserving memory of initial conditions. If the unitarity of the time evolution operator is broken, however, a dominant eigenvalue can appear in the spectrum, and the system can lose its sensitivity to the initial conditions. 

In this work we show how a system evolving under a fixed non-unitary evolution operator can remain sensitive to initial conditions at late times, and determine how this sensitivity is encoded in spectral statistics. Unlike those relevant to physical open systems, the non-unitary evolution operators studied here transform pure states into pure states, and so can be viewed as describing the contraction of translation-invariant tensor networks, or post-selected monitored dynamics. In our models, qubits evolve under the alternating action of a random unitary operator and a non-unitary field. The field biases the qubits towards a fixed product state, and the random unitary operators are nonlocal and fixed for all time. To probe the system's sensitivity to its initial conditions, we study the purification of mixed initial states, and a central question is how the purification dynamics is related to spectral properties of the evolution operator.

This relation can be understood, in part, via Yamamoto's theorem \cite{Yamamoto1967}, which connects the eigenvalues of our evolution operators $T$ to the singular values of $T^t$ in the limit of large (integer) time $t$. The squared singular values of $T^t$ are themselves eigenvalues of the time-evolved maximally mixed state, and so characterize purification. First we show that the time scale associated with purification is exponentially large in the number of qubits. This suggests that the gap between the magnitudes of leading eigenvalues is exponentially small. To develop insight into the eigenvalue magnitudes, we analyze the effect of an infinitesimal variation of the evolution operator, showing that the repulsion between eigenvalues along the azimuthal direction in the complex plane, which is characteristic of random unitary operators, is accompanied by radial \emph{eigenvalue attraction} when $T$ is non-unitary. 

Drawing on exact results on asymptotic properties of large non-unitary random matrices \cite{Wei2008,Bogomolny2010}, we then calculate the density of eigenvalues in the complex plane. We find that this distribution has sharp edges at large radii, reminiscent of Ginibre matrices \cite{Ginibre1965}, and that the eigenvalue density remains exponentially large near these edges. This high eigenvalue density is associated with an exponentially small gap between magnitudes of leading eigenvalues, and we numerically verify the relation between this eigenvalue gap and the exponentially large purification time. 

The slow purification that we identify implies that a generic initial pure state continues to explore an exponentially large space even at times that are themselves exponentially large. In unitary systems, the exploration of a large Hilbert space can be related, through generalizations \cite{Kos2018,chan2018spectral,bertini2018exact,Garratt2021} of Berry's diagonal approximation \cite{Berry1985}, to the emergence of spectral statistics resembling those of random matrices. We show that such a relation survives in the non-unitary setting through an analytical calculation of the (ensemble-averaged) spectral form factor (SFF). We find that azimuthal eigenvalue repulsion survives close to the (exponentially dense) outer edge of the eigenvalue distribution, and provide an interpretation of the different contributions to the SFF in terms of the interference between paths in Hilbert space. Notably, the averaged SFF exhibits a characteristic ramp behavior up to a time scale of the order of the purification time.

Our work builds on the study of purification transitions in the context of monitored many-body quantum dynamics \cite{Gullans2020,Choi2020}. There, it has been  shown that, under dynamics that is both non-unitary and random in time, two dynamical phases can exist. If the breaking of unitarity is weak, the system can remain sensitive to its initial conditions up to times that are exponential in the number of qubits. Solvable models for this regime involving global Haar random unitary operators were recently introduced in Refs.~\cite{Bulchandani2024,DeLuca2024}. In those models it was shown analytically that the sensitivity of the system to its initial conditions is encoded in an exponentially small gap in the set of Lyapunov exponents characterizing the singular values of the evolution operator, which is a product of random matrices (see also Refs.~\cite{Zabalo2022,Kumar2024,Aziz2024,Mochizuki2024,Mochizuki2025} for numerical studies of Lyapunov spectra in spatially structured non-unitary systems). By contrast, here our focus is on dynamics generated by an evolution operator that is the same for all time steps.

One motivation for our work comes from the study of computational complexity through the lens of tensor networks \cite{verstraete2006criticality,schuch2007computational,gonzalez2024random,chen2025sign,jiang2025positive}. It is known that the solutions to wide varieties of computational problems can be encoded in tensor networks \cite{verstraete2006criticality,schuch2007computational}, notable examples being the determination of ground state energies of certain two-dimensional statistical mechanics models, which is NP-hard in the worst case \cite{barahona1982computational}, and the simulation of arbitrary post-selected quantum computations, which would allow for the solution of PP-complete problems \cite{aaronson2005quantum}. Under the widely held assumption that classical computers cannot efficiently solve such problems, tensor network contraction must be extremely costly in the worst case. This contraction can be performed via multiplication of transfer matrices defining the tensor network and, when these transfer matrices are unitary, complexity can arise from the fact that `distant' parts of the network are highly correlated with one another. The weakly non-unitary evolution operators that we study in this work are models for the behavior of transfer matrices close to this regime. Here, distant parts of the corresponding tensor network are highly correlated because the non-unitary transfer matrix has many eigenvalues of similar magnitude.

The rest of the manuscript is organized as follows.
In Sec.~\ref{sec:model} we introduce our non-unitary evolution operator, and identify the time scale at which the system loses its sensitivity to its initial conditions. In Sec.~\ref{sec:purify} we compute the purification dynamics at times earlier than this characteristic scale and compare to it the behavior in models whose evolution operators vary randomly in time.
In Sec.~\ref{sec:radial} we relate the slow purification in our model to the spectrum of the evolution operator. First we identify radial level attraction as a mechanism leading to a small gap between eigenvalue magnitudes. Then, we analytically calculate the radial eigenvalue density and show that this gap is exponentially small, providing an explanation for slow purification. In Sec.~\ref{sec:SFF} we compute the SFF as a probe of azimuthal level repulsion, identifying another signature of chaotic dynamics that persists in spite of non-unitarity. We conclude with a discussion of further implications of the results in Sec.~\ref{sec:discussion}.

\section{Models and time scales}
\label{sec:model}
We consider a system of $N$ qubits $j=1,\ldots,N$, and so a Hilbert space with dimension $D = 2^N$. The qubits evolve under the alternating action of a \emph{fixed} Haar random $D \times D$ unitary operator $U$, and a set of single-qubit non-unitary fields $e^{h Z_j}$, where $Z_j$ is a Pauli matrix acting on qubit $j$. The evolution operator for a single time step is
\begin{align}
    T = \zeta U \equiv \Bigg[\bigotimes_{j=1}^{N} e^{h Z_j}\Bigg] U,
\end{align}
where $\zeta$ is the operator in square brackets on the right. The evolution over $t$ time steps, with $t$ integer, is $T(t) \equiv T^t$. The non-unitary field $h$ tends to push the qubits towards the eigenstates of $Z_j$ having unit eigenvalue, while $U$ delocalizes a typical state of the $N$ qubits over the $D$-dimensional Hilbert space. For $h=0$ the dynamics is unitary. Averages of $T$-dependent quantities over the ensemble of Haar random $U$ are indicated with angular brackets $\braket{\cdots}$. Since the evolution operator for each time step is the same, we say that the dynamics has discrete time-translation symmetry.

It will be useful to compare the dynamics described by $T(t)=T^t$ with a related operator, $\tilde T(t)$, which does not have time-translation symmetry but which otherwise has the same structure. We define this operator as
\begin{align}
	\tilde T(t) = \zeta U_t \zeta U_{t-1} \cdots \zeta U_1,
\end{align}
where $U_l$ with $l=1,\cdots,t$ are statistically independent Haar-random $D \times D$ unitary matrices. 

We will be interested in two different decompositions of the evolution operator $T(t)$. The spectral decomposition of $T$ has the form
\begin{align}
    T = \sum_{\alpha=0}^{D-1} \lambda_{\alpha}\ket{r_{\alpha}}\bra{l_{\alpha}},\label{eq:1}
\end{align}
and we write $\lambda_{\alpha} = \exp\big[\rho_{\alpha}+i\theta_{\alpha}\big]$ with $\rho_{\alpha} \geq \rho_{\alpha+1}$. For $h \neq 0$ the right and left eigenvectors, respectively $\ket{r_{\alpha}}$ and $\bra{l_{\alpha}}$, form a biorthonormal set $\braket{l_{\beta}|r_{\alpha}}=\delta_{\alpha \beta}$ rather than orthonormal sets: in general $\braket{r_{\beta}|r_{\alpha}} \neq \delta_{\alpha\beta}$. 

The singular value decomposition of $T(t)$, on the other hand, has the form
\begin{align}
    T(t) = \sum_{\alpha=0}^{D-1} \exp\big[ \sigma_{\alpha}(t)\big] \ket{u_{\alpha}(t)}\bra{v_{\alpha}(t)}. \label{eq:T}
\end{align}
Here the singular values $\exp[\sigma_{\alpha}(t)]$ as well as the left and right singular vectors, $\ket{u_{\alpha}(t)}$ and $\bra{v_{\alpha}(t)}$, respectively, depend on time $t$. The left and right singular vectors form independent orthonormal sets: $\braket{u_{\beta}(t)|u_{\alpha}(t)}=\delta_{\alpha \beta}$ and $\braket{v_{\beta}(t)|v_{\alpha}(t)}=\delta_{\alpha \beta}$. 

A key time scale in the dynamics generated by $T$, which will appear in several of our calculations below, is
\begin{align}
    t_*=D^{1/2}\Bigg[\frac{\mathbb{E}[\zeta^4]}{\mathbb{E}[\zeta^2]^2}-1\Bigg]^{-1/2},
\end{align}
where $\mathbb{E}[\zeta^x] \equiv D^{-1}\text{Tr}[\zeta^x]$. For finite ($N$-independent) $h$, at large $N$ we have $\mathbb{E}[\zeta^4]/\mathbb{E}[\zeta^2]^2-1 \approx \exp[b(h)N]$, with $b(h \to 0)=0$ and $b(h \to \infty)=\ln 2$. Therefore, the time scale
\begin{align}
	t_* \approx \exp\Bigg[\frac{N}{2}\Big(\ln 2 - b(h)\Big)\Bigg]\label{eq:tstarscaling}
\end{align}
is exponentially large in $N$. We will see that $t_*$ is related to the time scale over which a maximally mixed initial state purifies under the evolution operator $T(t)$. However, it is important to note that if we send $h \to 0$ at finite $N$ then this purification time necessarily diverges since the dynamics is then unitary. The scaling indicated in Eq.~\eqref{eq:tstarscaling} instead describes the behavior when $h$ is finite, and we increase the system size $N$. We will also see that the average $\langle K(t)\rangle$ of the spectral form factor $K(t)\equiv |\text{Tr}[T^t]|^2$ grows as $\mathbb{E}[\zeta^2]^t t$ until $t \sim t_*$. The factor of $t$ indicates random matrix level repulsion around the azimuthal direction in the complex plane.

\section{Slow purification}
\label{sec:purify}
Under unitary dynamics, two orthogonal quantum states will remain orthogonal for all time. Because orthogonal states can be perfectly distinguished, this means that information about the initial state is preserved for all time. Let us denote by $\{\ket{i}\}$ a complete orthonormal basis for the many-body Hilbert space with $i=1,\dots,D$, and by $\ket{i(t)}=T^t\ket{i}$ the (unnormalized) time-evolved state. In a non-unitary system we generically have $\braket{i(t)|j(t)}\neq 0$ for all $t \geq 1$. 

One way to probe the loss of orthogonality for time-evolved states, which will also allow us to establish connections to monitored quantum dynamics (in particular Refs.~\cite{DeLuca2024,Bulchandani2024}), is to study the purification of a maximally mixed initial density matrix $\Psi(0) = D^{-1}\sum_i \ket{i}\bra{i}$. After $t$ time steps the density matrix is
\begin{align}
	\Psi(t) = \frac{T^t \Psi(0) (T^t)^{\dag}}{\text{Tr}[T^t \Psi(0) (T^t)^{\dag}]},
\end{align}
and its purity can be expressed as 
\begin{align}
	\text{Tr}\Psi^2(t) = \frac{\sum_{i,j=1}^{D} |\braket{i(t)|j(t)}|^2}{\big(\sum_{i=1}^{D} |\braket{i(t)|i(t)}|\big)^2}. \label{eq:purity}
\end{align} 
Under unitary dynamics ($h=0$) we have $\text{Tr}[\Psi^2(t)]=D^{-1}$, while in a non-unitary system the development of nonzero overlaps $\braket{i(t)|j(t)}$ between time-evolved basis states with $i \neq j$ will cause the purity to increase. As $\ket{i(t)}/|\ket{i(t)}|$ approach one another, the system purifies, and all memory of the initial conditions is lost.

Purification is controlled by the singular values $\exp\big[ \sigma_{\alpha}(t)\big]$ of the evolution operator $T^t$. This follows from the fact that the spectral decomposition of the density matrix $\Psi(t)$ is 
\begin{align}
	\Psi(t) = \frac{1}{D} \sum_{\alpha=0}^{D-1} \exp\big[ \sigma_{\alpha}(t)\big] \ket{u_{\alpha}(t)}\bra{u_{\alpha}(t)}.
\end{align}
Quantities such as the purity $\text{Tr}\Psi^2(t)$, the von Neumann entropy $S(t) = S_1(t) = -\text{Tr}[\Psi(t)\ln \Psi(t)]$, and the R\'{e}nyi entropies $S_n(t) = [1-n]^{-1}\ln \text{Tr}\Psi^n(t)$ are all functions of the singular values $\exp\big[ \sigma_{\alpha}(t)\big]$ of $\Psi(t)$. 

A central quantity is the time scale $t_P$ for purification, which we define as the time beyond which $\braket{S_n(t)} \ll 1$ for $n \neq 0$. The behavior of $\braket{S_n(t)}$ at early times~(long before the purification time) can be determined using the Weingarten calculus \cite{samuel1980integrals,brouwer1996diagrammatic,collins2003moments} combined with a replica trick. The $2M^{\text{th}}$ moment over the Haar ensemble of unitary operators is given by 
\begin{align}
	\langle \prod_{m=0}^{M-1} U_{i_m j_m} U^*_{i^*_m j^*_m} \rangle = &\sum_{\sigma \tau} \text{Wg}(\sigma \tau^{-1})\label{eq:Weingarten} \\ &\times \prod_{m=0}^{M-1} \delta_{i_m i^*_{\sigma(m)}}\delta_{j_m j^*_{\tau(m)}}.\notag
\end{align}
The sums on the right are over permutations $\sigma$ and $\tau$ in the symmetric group $\mathcal{S}_M$ on $M$ elements, and $\text{Wg}(\sigma \tau^{-1})$ is a Weingarten function. For large $D$, the Weingarten functions scale as $\text{Wg}(\sigma \tau^{-1}) \sim D^{-M-|\sigma \tau^{-1}|}$, where $|\sigma \tau^{-1}|$ is the smallest number of pairwise transpositions that is equal to $\sigma \tau^{-1}$. Note that averages involving different numbers of $U$ and $U^*$ vanish. We will focus on the behavior of the second R\'{e}nyi entropy $\braket{S_2(t)}$, and to calculate this quantity we rely on the standard replica trick~\cite{vasseur2019entanglement}
\begin{align}
	\braket{S_2(t)} = \lim_{k\to0}\frac{1}{k}\bigg[&\langle\text{Tr}[T(t)T^\dagger(t)]^{2k}\rangle \label{eq:S2main} \\ &-\langle\text{Tr}[(T(t)T^\dagger(t))^2]^k\rangle\bigg].\notag
\end{align}
To guide our calculation of $\braket{S_2(t)}$ for evolution described by $T(t)$, we will first review the (simpler) calculation of the averaged second R\'{e}nyi entropy for $\tilde T(t)$, the evolution operator without time-translation invariance. We denote this quantity by $\braket{\tilde S_2(t)}$, and we evaluate it by replacing $T(t)$ in Eq.~\eqref{eq:S2main} with $\tilde T(t)$.

\begin{figure*}[t!]
\centering{}
\includegraphics[width=\textwidth]{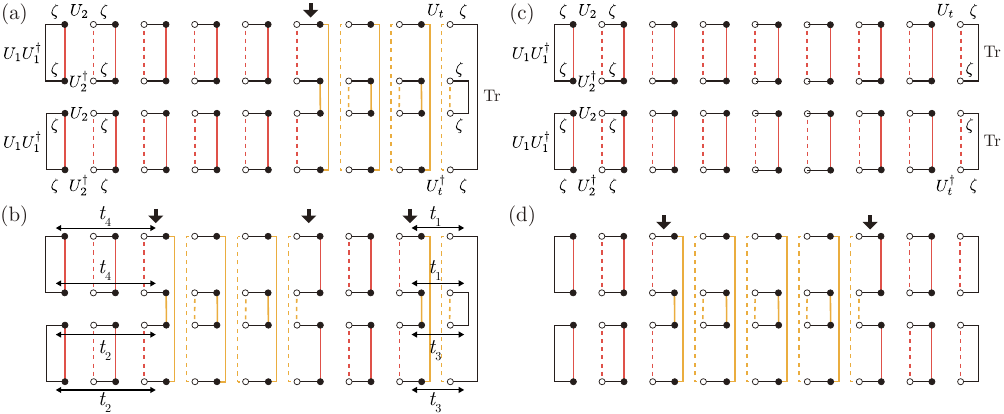}
\caption{\label{fig:Random}
Calculations of the averages (a,b) $\langle\text{Tr}[(\tilde{T}(t)\tilde{T}^\dagger(t))^2]\rangle$ and (c,d)~$\langle\text{Tr}[\tilde{T}(t)\tilde{T}^\dagger(t)]^2\rangle$ using Eq.~\eqref{eq:Weingarten}. Here we show the dominant pairs of permutations, which have $\sigma=\tau$. In each diagram, $\circ$ and $\bullet$ represent basis states $i_m,i_m^*$ and $j_m,j_m^*$ (corresponding to row and column indices, respectively, of $U$ and $U^*$) in Eq.~(\ref{eq:Weingarten}), and each horizontal bond corresponds to $\zeta$. Solid and dashed vertical lines represent the permutations $\sigma$ and $\tau$, with red and yellow indicating the identity and two-replica swap, respectively. Domain walls, indicated by thick arrows, appear when the permutations change across a layer of $\zeta$, leading to higher moments $\text{Tr}[\zeta^{2n}]$ with $n>1$.
(a,b)~Diagrams with one~(a) and three~(b) domain walls. After summing over the allowed locations of domain walls, the overall contributions from these classes of diagrams are $t\text{Wg}(1^2)^{t-1}\text{Tr}[\zeta^2]^{2t-2}\text{Tr}[\zeta^4]$ and $\frac{t^3}{6}\text{Wg}(1^2)^{t-1}\text{Tr}[\zeta^2]^{2t-6}\text{Tr}[\zeta^4]^3$, respectively.
(c,d)~Diagrams with zero~(c) and two~(d) domain walls. The overall contributions from these classes of diagrams are $\text{Wg}(1^2)^{t-1}\text{Tr}[\zeta^2]^{2t}$ and $\frac{t^2}{2}\text{Wg}(1^2)^{t-1}\text{Tr}[\zeta^2]^{2t-4}\text{Tr}[\zeta^4]^2$, respectively.
}
\end{figure*}

\subsection{Purification without time-translation invariance} 
The purification of a maximally mixed initial state under $\tilde T(t)$ takes an amount of time that is exponential in system size. To see this we calculate each of $\langle\text{Tr}[\tilde T(t)\tilde T^\dagger(t)]^{2k}\rangle$ and $-\langle\text{Tr}[(\tilde T(t)\tilde T^\dagger(t))^2]^k\rangle$ using Eq.~\eqref{eq:Weingarten}. In the absence of time-translation invariance, the averages can be performed independently for different time steps, with each determined by Eq.~\eqref{eq:Weingarten} with $M=2k$. Performing the averages associates each time step with a pair of permutations $\sigma,\tau \in \mathcal{S}_{2k}$, so each of the terms contributing to $\braket{S_2(t)}$ before we take the $k \to 0$ limit becomes a sum over configurations of permutations. Configurations where the permutations change in time, typically referred to as domain walls~\cite{DeLuca2024}~[Fig.~\ref{fig:Random}], are suppressed by powers of $D$. Crucially, for large $D$ these domain walls are well-separated in time, so can be viewed as independent of one another. 

Additional simplifications occur for $t\ll t_*^2$. First, the averages $\langle \text{Tr} [(\tilde T(t)\tilde T^{\dagger}(t))^2]\rangle$ and $\langle\text{Tr}[(\tilde T(t)\tilde T^{\dagger}(t))]^2\rangle$ are respectively dominated by permutations with one and zero domain walls. For $\langle \text{Tr} [(\tilde T(t)\tilde T^{\dagger}(t))^2]\rangle$, two operator copies of $\tilde T(t)\tilde T^{\dagger}(t)$ are connected by identity and swap permutations at the boundaries~[Fig.~\ref{fig:Random}(a,b)]. At each time step, the identity permutation connects each $\tilde{T}(t)$ with its own conjugate $\tilde{T}^\dagger(t)$ within the same copy, whereas the swap connects $\tilde{T}(t)$ in one copy with $\tilde{T}^\dagger(t)$ in another. When restricting to permutations with $\sigma=\tau$ at all time steps, the number of domain walls is always odd. Collecting the contributions from $\sigma=\tau$ with one and three domain walls, together with the lowest order corrections from $\sigma\neq\tau$~(see Appendix~\ref{sec:appA1}), gives
\begin{align}
\langle \text{Tr} [(\tilde T(t)\tilde T^{\dagger}(t))^2]\rangle=D\mathbb{E}[\zeta^2]^{2t}\bigg[1+\frac{Dt}{t_*^2}+\frac{Dt^3}{6t_*^6}+\cdots\bigg].\label{eq:domainmain}
\end{align}
The factors of $t$ and $t^3/6$ in the second and the third terms in Eq.~\eqref{eq:domainmain} correspond to the number of locations of one and three domain walls, respectively. This result indicates that permutations acting on operator copies that generate additional domain walls are negligible for $t\ll t_*^2$.

Similarly, for $\langle \text{Tr} [\tilde T(t)\tilde T^{\dagger}(t)]^2\rangle$, the permutations at the boundary are both identities~[Fig.~\ref{fig:Random}(c,d)], such that the number of domain walls is always even when restricting to $\sigma=\tau$. We then have
\begin{align}
\langle \text{Tr} [\tilde T(t)\tilde T^{\dagger}(t)]^2\rangle=D^2\mathbb{E}[\zeta^2]^{2t}\bigg[1+\frac{t^2}{2t_*^4}+\cdots\bigg],
\end{align}
where the factors of $1$ and $t^2/2$ arise from zero and two domain walls, respectively. This implies that $\langle\text{Tr} [\tilde T(t)\tilde T^{\dagger}(t)]^2\rangle\approx\langle\text{Tr} [\tilde T(t)\tilde T^{\dagger}(t)]\rangle^2\approx D^2\mathbb{E}[\zeta^2]^{2t}$, suggesting permutations that couple $\tilde{T}(t)$ and $\tilde{T}^\dagger(t)$ belonging to different copies of traces are negligible for $t\ll t_*^2$. Therefore, $\langle\text{Tr}[(\tilde T(t)\tilde T^{\dagger}(t))^2]^{k}\rangle$ is dominated by a set of `disconnected' permutations, and as a result
\begin{align}
	\langle\text{Tr}[(\tilde T(t)\tilde T^{\dagger}(t))^2]^k\rangle \approx \langle\text{Tr}[(\tilde T(t)\tilde T^{\dagger}(t))^2]\rangle^k,\label{eq:S2term1}
\end{align} 
and similarly,
\begin{align}
	\langle\text{Tr} [\tilde T(t)\tilde T^{\dagger}(t)]^{2k} \rangle \approx \langle \text{Tr} [\tilde T(t)\tilde T^{\dagger}(t)]\rangle^{2k},\label{eq:S2term2}
\end{align}
for $t\ll t_*^2$ up to corrections of order $t^2/t_*^4$. Combining these expressions and taking the $k \to 0$ limit we find 
\begin{align}
	\langle \tilde S_2(t)\rangle \approx \ln D - \ln \Big[ 1 + Dt/t_*^2 \Big], \quad t \ll t_*^2, \label{eq:S2tilde}
\end{align}
provided $D \gg 1$. As $t$ approaches $t_*^2$ the second term approaches $\ln D$, so the purification time $t_P$ is at least of order $t_*^2$. Beyond $t_*^2$ the purification is controlled solely by two leading singular values, and it can be shown that the latest times $\langle \tilde S_2(t)\rangle$ decays exponentially to zero with a characteristic time scale of order $t_*^2$~(see Appendix~\ref{sec:appA3}).

\subsection{Purification with time-translation invariance} 
At early times the decay of the averaged second R\'{e}nyi entropy in the model $T(t)=T^t$ coincides with that in $\tilde T(t)$. An important difference in the calculation of $\langle S_2(t)\rangle$ compared with $\langle \tilde S_2(t)\rangle$ is that the random unitary operations at different time steps are no longer independent, and so the expressions for $\langle\text{Tr}[(T^t(T^\dagger)^t)^2]^k\rangle$ and $\langle\text{Tr}[T^t(T^\dagger)^t]^{2k}\rangle$ in terms of Weingarten functions in principle involve sums over permutations $\sigma,\tau \in \mathcal{S}_{2tk}$. As a result, the subleading corrections become significant at a parametrically shorter time $t_*$, and the averages factorize over different time steps only at times $t \ll t_*$.

\begin{figure}[t!]
\centering{}
\includegraphics[width=0.48\textwidth]{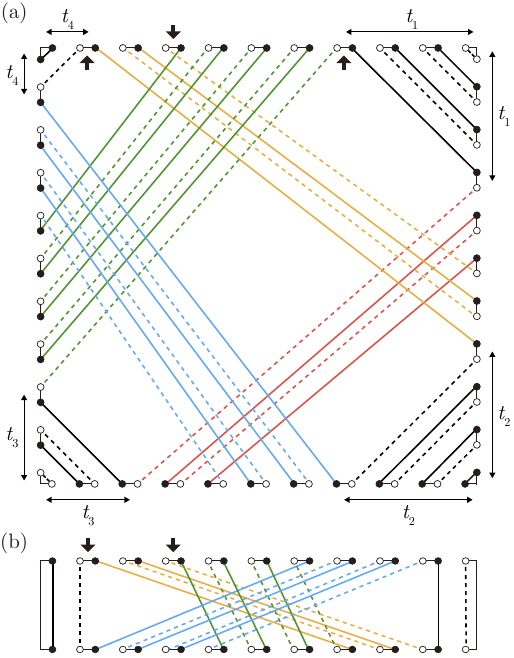}
\caption{\label{fig:TimeTI}
Schematics of next-to-leading pairs of permutations, with $\sigma=\tau$, involving non-equal-time pairings in the calculation of~(a)~$\langle\text{Tr}[(T^t(T^\dagger)^t)^2]\rangle$ and~(b)~$\langle\text{Tr}[T^t(T^\dagger)^t]\rangle$. Factors of $\text{Tr}[\zeta^4]$ arise from the configurations, indicated by thick arrows, where adjacent basis states~($\circ$ and $\bullet$ linked by a bond) are connected by distinct sets of equal-~(black) or non-equal-time~(colors other than black) pairings. In (a) the equal-time pairings over time intervals $t_1,t_2,t_3$ and $t_4$ are as shown in Fig.~\ref{fig:Random}(b) (although here the diagram is `unfolded'), while non-equal-time pairings generate three factors of $\text{Tr}[\zeta^4]$. The overall contribution from this class of diagrams, generated by summing over $t_1,\ldots,t_4$, is 
$\frac{t^5}{30}\text{Wg}(1^{2t-2})\text{Tr}[\zeta^2]^{2t-6}\text{Tr}[\zeta^4]^3$.
(b)~Non-equal-time pairings generate two factors of $\text{Tr}[\zeta^4]$.  The overall contribution from this class of diagrams is $\frac{t^4}{24}\text{Wg}(1^{t-1})\text{Tr}[\zeta^2]^{t-4}\text{Tr}[\zeta^4]^2$.
}
\end{figure}

To demonstrate this, we first analyze several leading contributions to $\langle\text{Tr}[(T^t(T^\dagger)^t)^2]\rangle$ and $\langle\text{Tr}[T^t(T^\dagger)^t]\rangle$. The dominant contributions coincide with those in the model $\tilde{T}(t)$, arising from equal-time pairings that are factorized over time steps with one and zero domain walls. However, time-translation invariance introduces additional non-equal-time pairings that connect $T$ and $T^\dagger$ at different time steps~[Fig.~\ref{fig:TimeTI}]. Configurations of these equal- and non-equal-time pairings can generate higher moments contributions beyond those captured by domain walls.

For $\langle\text{Tr}[(T^t(T^\dagger)^t)^2]\rangle$, the next-to-leading contribution arises from the diagram in Fig.~\ref{fig:TimeTI}(a) when restricting to $\sigma=\tau$, and is associated with three factors of $\text{Tr}[\zeta^4]$. This diagram can be viewed as the unfolded version of Fig.~\ref{fig:Random}(b), supplemented by additional non-equal-time pairings~(shown in colors). Denoting the lengths of equal-time-pairings at the corners as $t_{1,2,3,4}$, the factorized permutations $\sigma=\tau$ over time steps impose the constraints $t_1=t_3$ and $t_2=t_4$ in the model \emph{without} time-translation invariance~[Fig.~\ref{fig:Random}(b)]. Time-translation invariance lifts these constraints, i.e. we have sums over $t_{1,2,3,4}$, enhancing the combinatorial factor from $t^3/6$ to $t^5/30$~(see Appendix~\ref{sec:appA2}). Consequently,
\begin{align}
\langle\text{Tr}[(T^t(T^\dagger)^t)^2]\rangle=D\mathbb{E}[\zeta^2]^{2t}\bigg[1+\frac{Dt}{t_*^2}+\frac{Dt^5}{30t_*^6}+\cdots\bigg],
\end{align}
where the factor of $t$ corresponds to the same equal-time-pairing with one domain wall as in Eq.~(\ref{eq:domainmain}) and Fig.~\ref{fig:Random}(a).

A similar analysis applies for $\langle\text{Tr}[T^t(T^\dagger)^t]\rangle$. In the absence of time-translation invariance, $\langle\text{Tr}[\tilde{T}(t)\tilde{T}^\dagger(t)]\rangle\approx D\mathbb{E}[\zeta^2]^t$ is entirely given by the permutation factorized over time steps with zero domain wall. With time-translation invariance, the additional diagram shown in Fig.~\ref{fig:TimeTI}(b) contributes two factors of $\text{Tr}[\zeta^4]$. The number of diagrams in this class scales as $t^4/24$~(see Appendix~\ref{sec:appA2}), leading to
\begin{align}
\langle\text{Tr}[T^t(T^\dagger)^t]\rangle=D\mathbb{E}[\zeta^2]^t\bigg[1+\frac{t^4}{24t_*^4}+\cdots\bigg].
\end{align}
These expressions show that, although time-translation invariance modifies the behavior of $\langle\text{Tr}[(T^t(T^\dagger)^t)^2]\rangle$ and $\langle\text{Tr}[T^t(T^\dagger)^t]\rangle$, these modifications are negligible for $t\ll t_*$.

\begin{figure}[t!]
\centering{}
\includegraphics[width=0.48\textwidth]{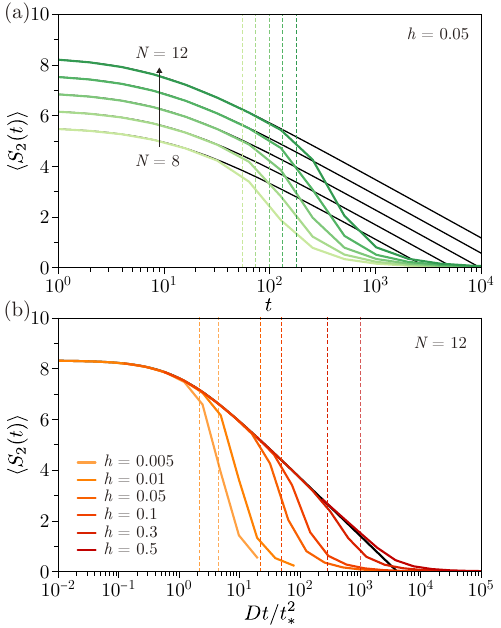}
\caption{\label{fig:Renyi2}
Purification of maximally mixed initial state under the model $T(t)=T^t$ with time-translation invariance.
(a)~Ensemble-averaged second R\'enyi entropy versus $t$ at several $N$ and $h=0.05$.
(b)~Ensemble-averaged second R\'enyi entropy versus $Dt/t_*^2$ at several $h$ and $N=12$.
In both panels, the black curves underneath represents early-time prediction of Eq.~(\ref{eq:EarlyEntropy}). The dashed lines indicate the time scales $t_*$ where we anticipate deviations from Eq.~(\ref{eq:EarlyEntropy}) due to time-translation invariance and, in panel (a), their equal separation on a logarithmic scale reflects the fact that $t_*$ is exponential in $N$.
}
\end{figure}

In addition, we show in Appendix~\ref{sec:appA2} that $\langle\text{Tr}[(T^t(T^\dagger)^t)^2]^k\rangle$ and $\langle\text{Tr}[T^t(T^\dagger)^t]^{2k}\rangle$ are dominated by a set of disconnected permutations for $t\ll t_*$, similar to Eqs.~(\ref{eq:S2term1}) and~(\ref{eq:S2term2}). Taking the $k\to 0$ limit we then have
\begin{align}
	\langle S_2(t)\rangle \approx \ln D - \ln \Big[ 1 + Dt/t_*^2 \Big], \quad t \ll t_*.\label{eq:EarlyEntropy}
\end{align}
This coincides with Eq.~\eqref{eq:S2tilde} except that it only applies over a much smaller~(but still exponentially large) window of times.

In Fig.~\ref{fig:Renyi2}, we numerically compute the ensemble-averaged second R\'enyi entropy at several non-unitary fields $h$ and system sizes $N$. There, we confirm that deviations from the early-time logarithmic decay in Eq.~(\ref{eq:EarlyEntropy}) become significant only for $t \gtrsim t_*$. As $t$ approaches $t_*$, using our expression for $t_*$ in Eq.~\eqref{eq:tstarscaling} it can be verified that $\langle S_2(t)\rangle$ remains extensive. Unfortunately, extending methods based on the Weingarten calculus beyond $t_*$ for the model with time-translation invariance becomes extremely complicated, as different time steps no longer approximately factorize in the average.

Nevertheless, the expression \eqref{eq:EarlyEntropy} for the entropy at $t \ll t_*$ shows that the purification time $t_P$ for the time-translation invariant model is at least $t_*$, and so remains exponential in system size. This raises the central question that we want to address: \emph{how is such slow purification encoded in the spectrum of the evolution operator $T$?}

It is useful to consider Yamamoto's theorem, which relates the singular values of powers of a matrix to its eigenvalues~\cite{Yamamoto1967}
\begin{align}
    \lim_{t \to \infty} \sigma_{\alpha}(t)/t = \rho_{\alpha}.\label{eq:Yamamoto}
\end{align}
In the simplest case, where the late-time state $\Psi(t)$ is dominated by contributions from just two singular values
\begin{align}
	\Psi(t) = &\big( 1 -e^{-[\sigma_0(t)-\sigma_1(t)]}\big) \ket{u_0(t)}\bra{u_0(t)}\\&+ e^{-[\sigma_0(t)-\sigma_1(t)]}\ket{u_1(t)}\bra{u_1(t)} + \cdots, \notag
\end{align}
we see that the purification time is related to the inverse of the (scaled) singular value gap $\Delta_{\sigma} = [\sigma_0(t)-\sigma_1(t)]/t$. One possibility is simply that this singular value gap approaches the radial eigenvalue gap $\Delta_{\rho} = \rho_0-\rho_1$ well before the purification time $t_P$. If this is the case, an exponentially long purification time must be associated with an exponentially small radial eigenvalue gap $\Delta_{\rho}$. 

\begin{figure}[t!]
\centering{}
\includegraphics[width=0.48\textwidth]{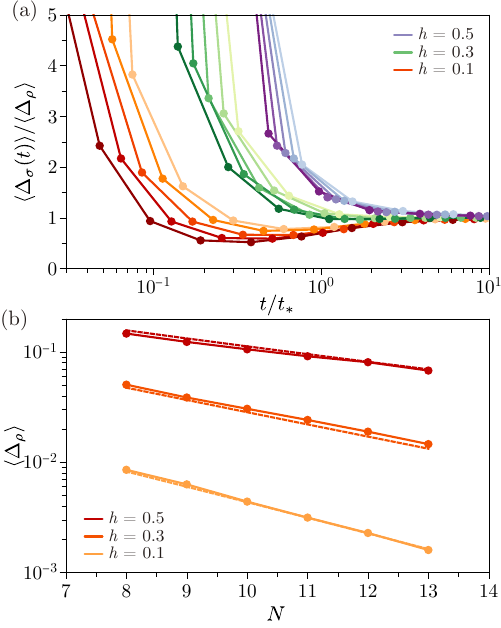}
\caption{\label{fig:Gap}
(a)~Ratio of averages $\braket{\Delta_\sigma(t)}/\braket{\Delta_\rho}$, with $\Delta_{\sigma}\equiv t^{-1}[\sigma_0(t)-\sigma_1(t)]$ and $\Delta_{\rho}=\rho_0-\rho_1$, as a function of $t/t_*$ for various system sizes $N=8,9,10,11,12$~(increasing opacity) and values of $h$. As $t$ increases and exceeds $t_*$, the singular value gap converges to the eigenvalue gap, as required by Yamamoto's theorem.
(b)~The eigenspectral gap of Eq.~(\ref{eq:1}) is exponentially small in $N$. The dashed lines show $\langle\Delta_\rho\rangle =  A t_*^{-1}\big[\ln t_*\big]^{-1/2}$ [Eq.~(\ref{eq:Gap})] with the coefficient $A \approx 0.4$.}
\end{figure}

In Fig.~\ref{fig:Gap} we present numerical evidence that our model exhibits this behavior. First, in Fig.~\ref{fig:Gap}(a), we show that the average $\langle\Delta_\sigma\rangle = t^{-1}\langle\sigma_0(t)-\sigma_1(t)\rangle$ converges to the average $\langle\Delta_\rho\rangle = \langle\rho_0-\rho_1\rangle$ at $t \sim t_*$. However, from Eq.~\eqref{eq:EarlyEntropy} it is clear that the entropy is still extensive at this time. This suggests that the time scale for the entropy to fall to well below unity (the purification time) is set by the inverse $\Delta_{\rho}^{-1}$ of the gap between the magnitudes of the leading eigenvalues. In Fig.~\ref{fig:Gap}(b), we show that the average $\langle\Delta_\rho\rangle$ of this gap over the ensemble of $T$ is exponentially small in $N$. 

\section{Radial eigenvalue statistics}
\label{sec:radial}
The behavior described in the previous section shows that slow purification is associated with small radial gaps $\rho_{\alpha}-\rho_{\beta}$. Moreover, even at the exponentially long time scale $t_*$, we have seen that the average of the second R\'{e}nyi entropy $S_2(t)=-\ln \text{Tr}\Psi^2(t)$ is extensive, suggesting that exponentially many singular values contribute to $\Psi(t)$ at these large times. Given Eq.~\eqref{eq:Yamamoto}, one possibility is therefore that the radial density of eigenvalues remains exponentially large in $N$ even close to the outer edge of the distribution of eigenvalues in the complex plane. 

This section focuses on the distribution of $\rho_{\alpha}$, and on the relation between this distribution and the purification time. In Sec.~\ref{sec:attraction} we identify a simple mechanism that can lead to small gaps $\rho_{\alpha}-\rho_{\beta}$. Then, in Sec.~\ref{sec:radialdensity} we leverage results from free probability theory to determine the radial eigenvalue density. Finally, in Sec.~\ref{sec:purificationtime} we relate the radial eigenvalue density of $T$ to the purification time under $T(t)=T^t$.

\subsection{Radial eigenvalue attraction}\label{sec:attraction}
Although the behavior identified above relates the purification time to the eigenvalue spectrum, it is \emph{a priori} unclear why the spectrum should feature such a small gap $\Delta_{\rho}$. Recall that, in chaotic unitary systems (such as $T$ with $h=0$), small changes in the evolution operator cause the eigenvalues to `repel' around the unit circle. This phenomenon leads to the well-known spectral rigidity in unitary random matrices~\cite{dyson1962brownian,TracyWidom1993,Mehta2004}. For $h \neq 0$ the eigenvalues are no longer confined to the unit circle and, if they were to repel, one might expect that their freedom to move off into the complex plane would lead to a single eigenvalue dominating over all of the others.

The reason this does not happen is that, for $h \neq 0$, the azimuthal repulsion familiar from unitary systems is accompanied by \emph{radial eigenvalue attraction}. We can see how such attraction arises by considering an infinitesimal change in the unitary $U$ appearing in $T=\zeta U$, and by using second-order perturbation theory. Let us regard $T=\zeta U$ as one instance of a random matrix arising in the fictitious dynamics $T(s+d s)=T(s)e^{i H(s) ds}$, where Hermitian matrices $H(s)$ are drawn from the Gaussian unitary ensemble, $s$ parametrizes a continuous random walk through the ensemble of $T$ with fixed $\zeta$, and $\overline{H_{ij}(s) H^*_{kl}(s')} = \delta_{ik}\delta_{jl} \delta(s-s')$, where $\overline{\cdots}$ represents an average over random instances of $H(s)$. This defines a kind of Dyson Brownian motion \cite{dyson1962brownian} in the matrix ensemble, although here we will focus only on an infinitesimal change in $s$. 

\begin{figure}[t!]
\centering{}
\includegraphics[width=0.48\textwidth]{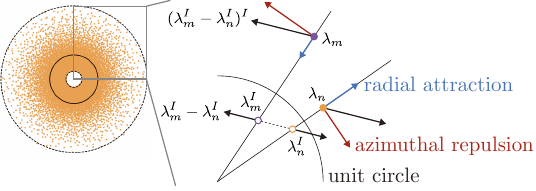}
\caption{\label{fig:DBM}
Left: eigenvalue distribution of $T=\zeta U$ within a ring whose outer and inner radii are respectively $e^{\pm\rho_N}=[\cosh^N(2h)]^{\pm1/2}$~(dashed).
The unit circle is shown solid.
Right: schematics of radial attraction between eigenvalues in non-unitary Dyson Brownian motion.
}
\end{figure}

Under such motion, the average change in the eigenvalue $\lambda_m=\lambda_m(s)$ from $s$ to $s+ds$ within second-order perturbation theory is
\begin{align}
	\overline{d\lambda_m} & =-\frac{2^N}{2}d s\lambda_m-d s\sum_{n\neq m}\frac{\lambda_m\lambda_n}{\lambda_m-\lambda_n}.\label{eq:DBM}
\end{align}
Viewing $\lambda_m$ as the position of the `particle' $m$ in the complex plane, we can interpret the right-hand side of Eq.~(\ref{eq:DBM}) as $F_m dt$ where $F_m = \sum_n F_{mn}$ is the total force acting on the particle \cite{dyson1962brownian} and $F_{mn}$ is the force between $\lambda_{m}$ and $\lambda_n$. This force can be expressed as 
\begin{align}
	F_{mn} = \frac{\lambda_m\lambda_n}{\lambda_n-\lambda_m}=(\lambda_m^{I}-\lambda_n^{I})^{I}
\end{align}
where the superscript $I$ denotes the circular inversion $z^I \equiv 1/z^*$. As shown in Fig.~\ref{fig:DBM}, the direction of this force is along the separation of their circular inversions $\lambda_m^I$ and $\lambda_n^I$. This results in the familiar azimuthal level repulsion, which for $h = 0$ can be understood as the origin of random matrix spectral statistics~\cite{dyson1962brownian}, as well as a new feature: radial level attraction. This radial attraction only arises for $h \neq 0$, since for $h = 0$ all eigenvalues have the same magnitude. Radial attraction is one possible mechanism allowing for the radial eigenvalue gap $\Delta_{\rho}$ to remain small even for the largest eigenvalues, and in the following we analyze the radial distribution in detail.

Although here we studied a single infinitesimal step in the motion of the eigenvalues, it is important to note that for $h \neq 0$ a complete description of their motion requires an analysis of their coupling to eigenvectors \cite{burda2014dysonian} (also see Appendix~\ref{sec:appB}). Only for $h = 0$ do eigenvalues and eigenvectors decouple. The existence of radial eigenvalue attraction nevertheless provides a useful heuristic understanding of spectral properties for small $h$. 

\subsection{Eigenvalue density}\label{sec:radialdensity}

Here we characterize the average density of eigenvalues of $T$ in the complex plane. For the ensemble of $T$ at fixed $h$, the ensemble-averaged density of eigenvalues is defined by
\begin{align}
	n(\rho,\theta) = \sum_{\alpha=0}^{D-1} \Big\langle\delta(\rho-\rho_{\alpha})\delta(\theta-\theta_{\alpha}) \Big\rangle,
\end{align}
where the average $\langle \cdots \rangle$ is over Haar random unitary operators $U$ appearing in $T$, and hence over the distributions of $\rho_{\alpha}$ and $\theta_{\alpha}$.

We first note that the eigenvalue distribution of $\zeta U$ is statistically invariant under independent left and right multiplication by Haar random unitary operators. Because of this invariance, the density of eigenvalues is rotationally invariant in the complex plane, $n(\rho,\theta)=n(\rho)/(2\pi)$, and the mean radial density $n(\rho)$ is completely determined by $\zeta$.

To make progress, we will rely on the exact formula for the ensemble-averaged radial eigenvalue distribution of $\zeta U$ derived in Refs.~\cite{Wei2008} and~\cite{Bogomolny2010}. Considering $0\leq\zeta_1\leq\cdots\leq\zeta_{D}$, the mean radial density $n(\rho)$ in the interval $\ln\zeta_{k}<\rho<\ln\zeta_{k+1}$ is there given by
\begin{align}
n(\rho)=\frac{D}{\pi i}\int_0^1 duw(u)\oint_{C_k}\frac{dv}{w(v)}\Big[D-\frac{u+v-1}{v-u}\Big],\label{eq:ExactDensity}
\end{align}
where $w(v)=\prod_{j=1}^{D}\big[v-(v-1)e^{-2\rho}\zeta_j^2\big]$ with $v$ complex, and the contour integral $C_k$ encloses poles of $w(v)^{-1}$ with $\text{Re}[v]>u$. At $D\gg1$, a continuous approximation to the density is obtained via the saddle point approach. Rewriting $w(v)=e^{D\Phi(v)}$, the saddle point $v=v_s$ of $\Phi(v)$ and $\rho$ satisfy the implicit relation
\begin{align}
\Phi^\prime(v_s)=\frac{1}{v_s}\bigg[1+\frac{1}{v_s-1}G_{\zeta^2}\Big(\frac{v_s-1}{v_se^{2\rho}}\Big)\bigg]=0,\label{eq:saddle}
\end{align}
with the moment-generating function of $\zeta^2$, $G_{\zeta^2}(u)=\frac{1}{D}\sum_{j=1}^D\frac{u\zeta_j^2}{1-u\zeta_j^2}$. The solution to Eq.~\eqref{eq:saddle} defines the function $v_{s}(\rho)$. The saddle point contribution to the integral over $v$ is then obtained by deforming the contour to pass through this saddle point. We note that, when $v_{s}(\rho)<u$, where $u\in[0,1]$ is the outer integration variable in Eq.~\eqref{eq:ExactDensity}, there is an additional contribution to the integral over $v$ from the pole $v=u$. This contribution is $-2D\int_{v_s}^1du(2u-1)=2Dv_s(v_s-1)$ at $0<v_s<1$. Combining these two contributions, the large $D$ asymptotic of Eq.~(\ref{eq:ExactDensity}) reads \cite{Wei2008,Bogomolny2010}
\begin{align}
n&(\rho)=2Dv_s(v_s-1)\Theta(v_s)\Theta(1-v_s)\label{eq:ContDensity}
\\
& +\frac{D}{|\Phi^{\prime\prime}|}\bigg(\text{Erf}\Big[\sqrt{\frac{D|\Phi^{\prime\prime}|}{2}}(1-v_s)\Big]+\text{Erf}\Big[\sqrt{\frac{D|\Phi^{\prime\prime}|}{2}}v_s\Big]\bigg),\nonumber
\end{align}
where in this expression we have written $\Phi^{\prime\prime}=\Phi^{\prime\prime}(v_s)$ for brevity.

We will use these results to investigate the behavior $n(\rho)$ at large $\rho$ in our model with $\zeta=e^{h\sum_{j=1}^NZ_j}$. It is evident that Eq.~\eqref{eq:ContDensity} separates two regimes: $0<v_s(\rho)<1$, and $v_s(\rho)$ outside this interval. We refer to the former as the `bulk' region, as this contains the majority of the eigenvalues, and to the latter as the tail. In our model, in the tail the density of eigenvalues is exponentially suppressed in $N$. 

From the saddle point condition Eq.~(\ref{eq:saddle}) we find that the bulk corresponds to $\rho\in [-\rho_N,\rho_N]$, where 
\begin{align}
    \rho_N = \frac{1}{2}\ln \mathbb{E}[\zeta^2] = \frac{N}{2}\ln \cosh(2h).
\end{align}
At the outer bulk edge $\rho=\rho_N$ the mean radial density is exponentially large in $N$, and is given by
\begin{align}
    n(\rho_N)=t_*^2 \approx \big[1+\text{sech}(4h)\big]^N.\label{eq:DOS}
\end{align}
For $\rho > \rho_N$, the mean radial density $n(\rho)$ decays rapidly with increasing $\rho-\rho_N$, at a rate that is \emph{exponential} in $N$, i.e. $n(\rho)$ has sharp edges at large $N$. In Appendix~\ref{sec:appC} we derive the form of the tail explicitly; for large $N$ and $\rho > \rho_N$ we find
\begin{align}
    \frac{n(\rho)}{n(\rho_N)} = 1-\text{Erf}\Big[\sqrt{2}t_*\rho_N\big(\rho/\rho_N-1\big)\Big].\label{eq:Tail}
\end{align}
Crucially, although the tail beyond $\rho_N$ decays exponentially with $\rho-\rho_N$, over a scale $1/t_*$, the massive mean radial density at the bulk edge ensures that the number $\mathcal{N}$ of eigenvalues in this tail is still exponentially large in $N$:
\begin{align}
    \mathcal{N} \equiv \int_{\rho>\rho_N}n(\rho)d\rho= t_*/\sqrt{2\pi}.
\end{align}
Because of this, we will find that this tail controls purification dynamics at late times $t \gtrsim t_*$. 

\begin{figure}[t!]
\centering{}
\includegraphics[width=0.48\textwidth]{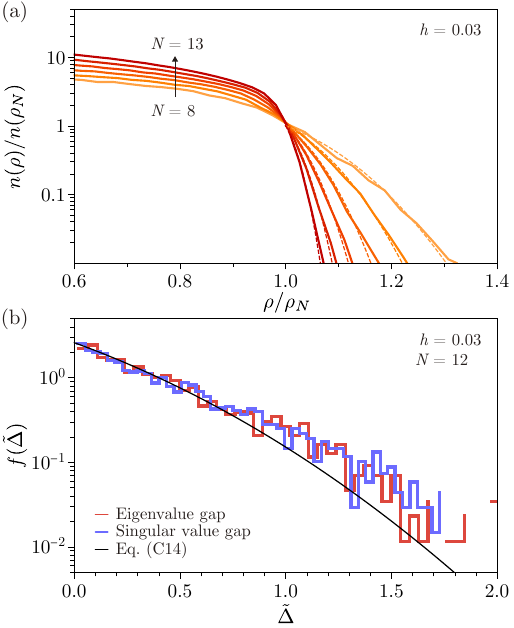}
\caption{\label{fig:DOS}
(a)~Scaled mean radial eigenvalue density around the outer bulk edge at $h=0.3$. The results are obtained from $2000$ realizations for each $N$. The dashed lines follow Eq.~(\ref{eq:Tail}) and agree well with numerics.
(b)~Scaled probability density~($\tilde{\Delta}=\sqrt{2n(\rho_N)}\Delta$) of leading eigenvalue gap $\Delta_\rho$, leading singular value gap $\Delta_\sigma$ at $t=10t_*$, and theoretical prediction in Eq.~(\ref{eq:LeadingGap}) at $N=12$, $h=0.3$.
}
\end{figure}

In Fig.~\ref{fig:DOS}(a) we compare the true radial density $n(\rho)$, computed via exact diagonalization in finite systems, with the theoretical prediction Eq.~\eqref{eq:Tail}. Rescaling $n(\rho)$ by the analytical prediction for $n(\rho_N)$, which applies at large $N$, we see a clear convergence of $n(\rho)/n(\rho_N)$ to a sharp-edged function of the ratio $\rho/\rho_N$, which is described by Eq.~(\ref{eq:Tail}) and shown in Fig.~\ref{fig:DOS}(a) in dashed lines.

Given the structure of the tail $n(\rho>\rho_N)$, we can estimate the probability distribution $f(\Delta_{\rho})$ of the leading eigenvalue gap $\Delta_{\rho}=\rho_0-\rho_1$ by assuming that the magnitudes of the eigenvalues with $\rho > \rho_N$ are independently and identically distributed. The result is a Poisson distribution $f(\Delta_\rho)\sim \langle\Delta_\rho\rangle^{-1} e^{-\Delta_\rho/\langle\Delta_\rho\rangle}$~(see Appendix~\ref{sec:appC4}), with characteristic scale given by the mean leading eigenvalue gap~[Fig.~\ref{fig:Gap}(b)]
\begin{align}
\langle\Delta_\rho\rangle \sim t_*^{-1}\big[\ln t_*\big]^{-1/2}.\label{eq:Gap}
\end{align}
In Fig.~\ref{fig:DOS}(b) we test our prediction for $f(\Delta_{\rho})$ numerically, and find good agreement in the regime $\Delta_{\rho} \lesssim \langle\Delta_\rho\rangle$. We can now discuss the implications of these eigenvalue statistics for the purification of mixed initial states.

\subsection{Purification time}\label{sec:purificationtime}
As we show in Eq.~(\ref{eq:EarlyEntropy}), the ensemble-averaged second R\'enyi entropy $\langle S_2(t)\rangle$ remains extensive up to an exponentially long time scale $t\sim t_*$ in the model $T(t)=T^t$ with time-translation invariance. We now relate this purification time scale to the spectrum of $T$.

At times $t$ of order $t_*$, the numerical results presented in Fig.~\ref{fig:Gap}(a) suggest that the leading singular values $\sigma_\alpha(t)$ approach their asymptotic values $t\rho_\alpha$ prescribed by Yamamoto's theorem. The implication is that the purification of $\Psi(t)$ is controlled by eigenvalues with radial component $\rho$ within width $\sim t_*^{-1}$ around the outer bulk edge $\rho_N$. As we have noted above, there are exponentially many eigenvalues within this window, which therefore make significant contributions to $\Psi(t)$ even at the time scale $t\sim t_*$. These contributions cause the R\'enyi entropies to remain extensive up to $t\sim t_*$.

\begin{figure}[t!]
\centering{}
\includegraphics[width=0.48\textwidth]{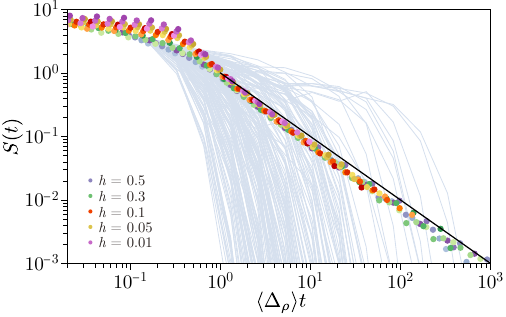}
\caption{\label{fig:Purify}
Scaling collapse of late-time ensemble-averaged von Neumann entropy~(points) versus $\langle\Delta_\rho\rangle t$. The theoretical prediction in Eq.~(\ref{eq:powerlaw}) is indicated by the black line. Increasing point opacity corresponds to $N=8,9,10,11,12$. The pale blue curves show the von Neumann entropy for individual realizations for $N=12, h=0.5$.
}
\end{figure}

At even longer times beyond the inverse of typical singular value gap, $t\gg \braket{\Delta_{\sigma}}^{-1}\approx\braket{\Delta_{\rho}}^{-1} \sim N^{1/2} t_*$, the evolution of $\Psi(t)$ is dominated by the two leading singular values. Hence, the von Neumann entropy is given by $S(t)\approx 2\Delta_\rho te^{-2\Delta_\rho t}$, where we used the fact that at these late times the singular values saturate to their asymptotic values prescribed by Yamamoto's theorem $\sigma_0(t)-\sigma_1(t) \approx t \Delta_{\rho}$. In Fig.~\ref{fig:DOS}(b), we numerically calculate the distribution of leading singular value gap at $t=10t_*$, which coincides with the distribution of leading eigenvalue gap $f(\Delta_{\rho})$. Using the approximate Poisson distribution $f(\Delta_{\rho})\sim \langle\Delta_\rho\rangle^{-1} e^{-\Delta_\rho/\langle\Delta_\rho\rangle}$ discussed in connection with Fig.~\ref{fig:DOS}(b), we find that the ensemble average of the entropy decays as a power law at these late times
\begin{align}
\langle S(t)\rangle\approx\int_{0}^{\infty}d\Delta_\rho f(\Delta_\rho)2\Delta_\rho te^{-2\Delta_\rho t}\approx \frac{1}{2\langle \Delta_\rho\rangle t},\label{eq:powerlaw}
\end{align}
The same power-law decay holds for all R\'{e}nyi entropies with $n>0$. In Fig.~\ref{fig:Purify}, we numerically calculate the ensemble-averaged von Neumann entropy at late times, which confirms the behavior predicted in Eq.~(\ref{eq:powerlaw}). We also observe large fluctuations in the late-time behavior between different realizations of $U$.

In contrast to this behavior, models with random time dependent $U$ lead to an exponential decay of R\'enyi entropies at the latest times (beyond a timescale $t_*^2$ that is exponential in $N$). This behavior can be understood in terms of the repulsion between the leading late-time singular values $\sigma_0(t)$ and $\sigma_1(t)$ \cite{Bulchandani2024}, which appears to be absent in the setting studied here.

\section{Azimuthal eigenvalue statistics}
\label{sec:SFF}
Our results so far show that, at time scales $t_*$ exponential in $N$, the number of singular values that make significant contributions to the state $\Psi(t) \propto T(t)T^{\dag}(t)$ is itself exponential in $N$. A generic initial pure state therefore has support on many eigenvectors at these late times, and hence the time-evolved state is highly sensitive to its initial conditions. This is a defining characteristic of dynamical chaos.

It is natural to ask whether a key signature of unitary quantum chaos, azimuthal level repulsion, survives at late times in our non-unitary model. By calculating the spectral form factor (SFF), $K(t)=|\text{Tr}[T^t]|^2=\sum_{m,n} e^{(\rho_m + \rho_n)t + i(\theta_m-\theta_n)t}$, we now show that azimuthal level repulsion indeed survives. In particular, we show that $\langle K(t)\rangle \propto t$ up to $t$ of order $t_*$.

\begin{figure}[t!]
\centering{}
\includegraphics[width=0.48\textwidth]{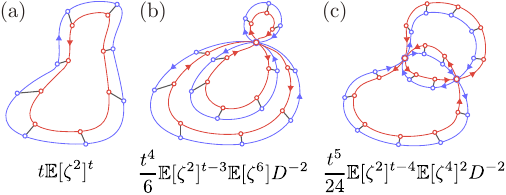}
\caption{\label{fig:Path}
Several leading pairings between forward~(blue) and backward~(red) paths and corresponding contributions to $\langle K(t)\rangle$ when $\sigma=\tau$.
(a)~Leading pairing with $t$ independent basis states.
(b)~One basis state is visited three times in each of the forward and backward paths.
(c)~Two basis are respectively visited twice in each of the forward and backward paths.
}
\end{figure}

By inserting resolutions of the identity into $\text{Tr}[T^t]$ we can view this trace as a sum over all closed paths in Hilbert space~(in the eigenbasis of $\zeta$)
\begin{align}
K(t)= & \sum_{a_0,\cdots,a_{t-1}}\zeta_{a_0}U_{a_0a_{t-1}}\cdots \zeta_{a_1}U_{a_1a_0}\label{eq:K}
\\
& \times \sum_{a_0^*,\cdots,a_{t-1}^*}\zeta_{a_0^*}U_{a_0^*a_{t-1}^*}^*\cdots \zeta_{a_1^*}U_{a_1^*a_0^*}^*,\nonumber
\end{align}
so that $K(t)$ is a discrete sum over all pairs of `forward' $(a_0a_1\dots a_{t-1})$ and `backward' $(a_0^*a_1^*\dots a_{t-1}^*)$ paths, contributing to $\text{Tr}[T^t]$ and its conjugate, respectively~\cite{Berry1985,Kos2018,chan2018spectral,Garratt2021}. The SFF is not self-averaging, and to wash out fluctuations we average over the ensemble of $U$. The ensemble-averaged SFF, $\langle K(t)\rangle$, can then be expressed as a sum over pairings of indices $a_r$ and $a_{r'}^*$
\begin{align}
\langle K(t)\rangle= & \sum_{\sigma\tau}\text{Wg}(\sigma\tau^{-1})\label{eq:avgSFF}
\\
& \times\sum_{\substack{a_0,\cdots,a_{t-1} \\ a_0^*,\cdots,a_{t-1}^*}}\prod_{r=0}^{t-1}\zeta_{a_{r+1}}\zeta_{a_{r+1}^*}\delta_{a_ra_{\sigma(r)}^*}\delta_{a_{r+1}a_{\tau(r+1)}^*}.\nonumber
\end{align}
For $h=0$, the dominant contribution comes from `diagonal' pairings $a_r=a_{r+s\text{\ mod\ }t}^*$, representing pairs of paths related by relative time-translation symmetry~[Fig.~\ref{fig:Path}(a)]. The independent time-translation invariances of forward and backward paths give rise to $t$ leading pairings, corresponding to $t$ cyclic permutations among $t$ indices, and thus $\langle K(t)\rangle=t$ at $h=0$. Retaining only these cyclic pairings of indices is analogous to the diagonal approximation in the semiclassical theory of quantum chaos \cite{Berry1985}. Generic paths explore the Hilbert space, and the complex amplitudes of typical pairs of paths are only weakly correlated with one another, so they do not survive an average. The only pairs of paths which survive are those whose amplitudes are related by symmetry, and these correspond to the cyclic pairings of indices. For $h=0$, and for $t \ll D$, the cyclic pairings alone capture the linear ramp $\langle K(t)\rangle = t$.

To analyze the impact of non-unitarity, we evaluate the corrections to the average SFF at $h \neq 0$. As time $t$ increases the SFF is now dominated by pairs of eigenvalues with increasing magnitudes $e^{\rho_m}$ and $e^{\rho_n}$, and the details of its time dependence will provide us with information on the statistics of phase differences $\theta_m-\theta_n$. For a pair of permutations $\sigma,\tau\in \mathcal{S}_{t}$ in Eq.~(\ref{eq:avgSFF}), with $\sigma \tau^{-1}$ consisting of $m$ cycles of lengths $n_k$, with $k=0,\ldots,(m-1)$ and $\sum_{k=0}^{m-1}n_k=t$, the summand in Eq.~(\ref{eq:avgSFF}) becomes a product over cycles, with a cycle of length $n_k$ contributing a factor $\text{Tr}[\zeta^{2n_k}]$. This gives the expression
\begin{align}
\langle K(t)\rangle=\sum_{\sigma\tau}\text{Wg}(\sigma\tau^{-1})\prod_{k=0}^{m-1}\text{Tr}[\zeta^{2n_k}].
\end{align}
Different cycle structures now appear with different weights, and it is useful to note that the cyclic permutations (with $m=t$ and $n_k=1$) appear with weight $\text{Tr}[\zeta^{2}]^t$. These permutations are associated with closed paths $(a_0,\ldots,a_{t-1})$ where all $a_r$ are independent, and as a consequence the trace structure involves $t$ independent sums over $D$ basis states. More generally, a cycle $k$ of length $n_k$ is associated with basis states that are revisited $n_k$ times, and the factor $\text{Tr}[\zeta^{2n_k}]$ means that contributions from basis states $a_r$ with large $\zeta_{a_r}$ are enhanced relative to those with small $\zeta_{a_r}$. As a result, the leading contribution $tD^{-t}\text{Tr}[\zeta^2]^t=t\mathbb{E}[\zeta^2]^t$ at $t\ll D$ stems from the sums over $t$ cyclic permutations $\sigma$, with $\tau=\sigma$, consistent with the diagonal approximation.

We note that all other contributions are suppressed by powers of $D$ for $h=0$ and $t\ll D$. This is because contributions from $\sigma\neq\tau$ and paths leading to $m<t$ independent indices are both suppressed by powers of $D$. However, for finite $h>0$ although the number of independent indices is reduced, there are large contributions from paths that revisit basis states with large $\zeta_{a_r}$, and these change the time-dependence of the ensemble-averaged SFF, allowing us to identify several subleading contributions as~(see Appendix~\ref{sec:appD})
\begin{align}
\langle K(t)\rangle= & \,\mathbb{E}[\zeta^2]^t\Bigg[t+\frac{t^4}{6}\bigg(\frac{\mathbb{E}[\zeta^6]}{\mathbb{E}[\zeta^2]^3}-3\frac{\mathbb{E}[\zeta^4]}{\mathbb{E}[\zeta^2]^2}+2\bigg)D^{-2}\nonumber
\\
& +\frac{t^5}{24}\bigg(\frac{\mathbb{E}[\zeta^4]}{\mathbb{E}[\zeta^2]^2}-1\bigg)^2D^{-2}+\mathcal{O}(D^{-4})\Bigg].\label{eq:TotalSFF}
\end{align}
Here the $t^4$ and $t^5$ terms arise from paths that visit one basis state six times and two basis states four times~[Fig.~\ref{fig:Path}(b,c)], respectively. By determining the time scales for these subleading terms to be comparable with leading contribution, and using $(D\mathbb{E}[\zeta^4])^{1/2}\geq(D\mathbb{E}[\zeta^6])^{1/3}$, we see that paths in Fig.~\ref{fig:Path}(c) contribute significantly to the SFF at a shorter time scale $t_*$ than those in Fig.~\ref{fig:Path}(b). Hence, for $h>0$ and $t\ll t_*$ we have
\begin{align}
\langle K(t)\rangle=\mathbb{E}[\zeta^2]^t t \bigg[ 1 +\frac{1}{24}\bigg(\frac{t}{t_*}\bigg)^4+\cdots\bigg],\label{eq:SFF}
\end{align}
where the cyclic permutations dominate for $t\ll t_*$, up to the time scale where the purification dynamics is controlled by the tail of eigenvalue distribution.

The exponential growth of the SFF as $\mathbb{E}[\zeta^2]^t$ for $t \ll t_*$ can be understood as a consequence of the fact that pairs of eigenvalues with $\rho \sim \ln \mathbb{E}[\zeta^2]^{1/2}=\rho_N$ dominate the average SFF, i.e. the growth is controlled by the outer edge of the eigenvalue distribution. To isolate the azimuthal level correlations, it is convenient to consider the scaled SFF, defined as a function of $s = t/t_*$ as
\begin{align}
 \kappa(s) = \mathbb{E}[\zeta^2]^{-t}t_*^{-1} K(t), \label{eq:kappadef}
\end{align}
where on the right the integer time $t = \lfloor s t_* \rfloor$. From Eq.~\eqref{eq:SFF} the ensemble average of the scaled SFF for $s \ll 1$ is
\begin{align}
    \langle \kappa(s)\rangle = s + \frac{1}{24}s^5 + \cdots. \label{eq:kappas}
\end{align}
Behavior of this kind was previously identified in a calculation of a scaled SFF for the Ginibre ensemble \cite{shivam2023many}. Here we have shown that Eq.~(\ref{eq:kappas}) emerges for the entire family of transfer matrices $T=\zeta U$. We have also provided an interpretation of the corrections to linear behavior in terms of paths in Hilbert space. We note also that, when $\zeta$ is a projector onto a subspace, $T=\zeta U$ corresponds to the truncated unitary ensemble, which is known to exhibit Ginibre spectral statistics \cite{Zyczkowski2000}. Our results show that Ginibre universality in the SFF applies beyond this hard truncation to a much broader class of operators.

For $s \gg 1$, where a small number of leading eigenvalues dominate, we approximate $\langle K(t)\rangle \approx \sum_{m=0}^{D-1}\langle e^{2\rho_m t}\rangle$. Then, using Eqs.~\eqref{eq:DOS} and \eqref{eq:Tail}, we find that $\langle \kappa(s)\rangle$ grows approximately exponentially with $s$. It is interesting to note that, when working at large $N$ and finite $h$, the SFF deviates from the ramp at the time scale $t_*$, which approaches $2^{N/2}$ as $h$ is decreased. On the other hand, if $h$ is sent to zero at finite $N$, then the SFF approaches the standard behavior for Haar random unitary matrices, i.e. $\langle K(t)\rangle =t$ for $1 \leq t \leq 2^N$ and $\langle K(t)\rangle = 2^N$ for $t \geq 2^N$. These behaviors indicate a non-commutation of the limits of weak non-unitarity and large system size. 

This behavior is most clearly seen by examining the average of the scaled SFF $\langle \kappa(s)\rangle$. After first sending $N \to \infty$ with $h \neq 0$, $\langle \kappa(s)\rangle$ is described by Eq.~\eqref{eq:kappas}, including when we subsequently take the limit $|h| \to 0$. In particular, $\langle \kappa(s)\rangle \approx s$ for $s \ll 1$. On the other hand, if we send $|h| \to 0$ at finite $N$, then we have $\langle \kappa(s) \rangle = s$ over a much wider window, $0 < s \leq 2^{N/2}$, corresponding to $0 < t < 2^N$. 

\begin{figure}[t!]
\centering{}
\includegraphics[width=0.48\textwidth]{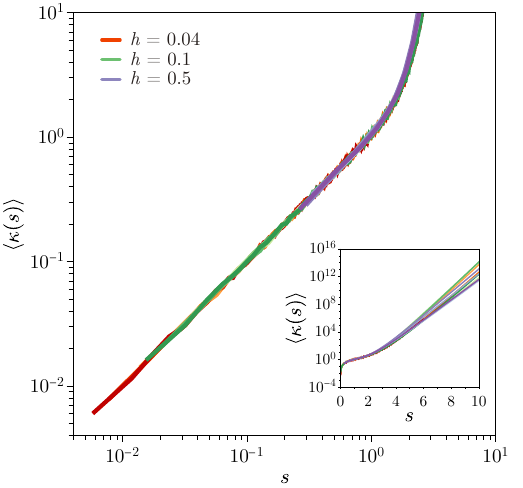}
\caption{\label{fig:SFF}
Scaling collapse of ensemble-averaged scaled SFF $\langle\kappa(s)\rangle$. Increasing line opacity corresponds to $N=8,9,10,11$.
Inset: Data in the main panel on log-linear scale, where the scaled SFF grows exponentially in time at $t\gtrsim t_*$.
}
\end{figure}

In Fig.~\ref{fig:SFF} we numerically calculate the average scaled SFFs $\langle \kappa(s)\rangle$ for various non-unitary fields $h$ and system sizes $N$. We find a clear collapse onto a single linear ramp up to $s$ of order unity, corresponding to an exponentially long time $t \sim t_*$, in agreement with Eq.~\eqref{eq:kappas}. The implication is that there is azimuthal level repulsion close to this outer edge, where the eigenvalue density is exponential in $N$. Beyond this time, azimuthal level repulsion is lost. These results show that the sensitivity to initial conditions, which is required for dynamical chaos, is lost at the same time scale at which we also lose this spectral signature of quantum chaos. 

\section{Discussion}
\label{sec:discussion}
In this work we have investigated the relation between two different notions of chaos in weakly non-unitary quantum systems. Under unitary classical dynamics, a defining feature of chaos is that the late-time state of a system is highly sensitive to initial conditions \cite{Gutzwiller1990}. This sensitivity can be visualized through the divergence between nearby trajectories in phase space. It is not straightforward to generalize this notion to quantum systems, where states evolve in Hilbert space rather than phase space, because under unitary quantum dynamics the overlaps between pure states do not change with time. 

Instead, the theory of quantum chaos is grounded in the analysis of spectral statistics \cite{bohigas1984characterization}. The connection between this idea, and the classical definition based on phase space trajectories, becomes sharp in the semiclassical limit. There it was recognized that when the classical analog of a quantum system is chaotic, in the semiclassical limit of that quantum system one finds spectral correlations resembling those of random matrices. In particular, the characteristic ramp in the SFF as a function of time can be associated with the ergodicity of classical phase paths \cite{Berry1985}. Crucially, the spectrum of a quantum system remains well-defined even when far from any semiclassical limit. For this reason, random matrix spectral statistics are often taken as a definition of quantum chaos. 

By introducing weak non-unitarity to an evolution operator describing the dynamics of pure quantum states, we arrive at a setting where these pure states can lose their sensitivity to their initial conditions. This loss of sensitivity can be characterized through the emergence of nonzero overlaps between initially orthogonal pure states and, therefore, by considering the purification of maximally mixed initial states under non-unitary dynamics. This behavior is encoded in the $t$-dependent singular values of the evolution operator for $t$ time steps which, at late times, approach powers of the magnitudes of eigenvalues. Through calculations of the entropy, we have shown that for finite non-unitary fields and for large system sizes the time scale for purification is exponentially large in the system size. This large purification time is associated with the existence of a sharp edge in the distribution of eigenvalues in the complex plane, around which the density of eigenvalue magnitudes is exponentially large in system size.

Even close to the sharp edge of the distribution of eigenvalue magnitudes, the correlations between phases of eigenvalues resemble those in random unitary matrices. Heuristically, this can be understood as arising from the combination of radial level attraction, which leads to a high radial eigenvalue density, and conventional azimuthal level repulsion. We demonstrated this feature of phase correlations by calculating the average SFF. After rescaling this quantity by powers of eigenvalue magnitudes, we showed that the ramp survives even close to the purification time. 

As in the unitary setting, this ramp can be associated with a diagonal approximation to a discrete sum over pairs of closed paths in Hilbert space \cite{Kos2018,chan2018spectral,bertini2018exact,Garratt2021}, and we have related the persistence of this ramp to a kind of ergodicity in individual paths. Although non-unitarity amplifies contributions to the SFF from paths that repeatedly revisit a subset of the basis states, it is only beyond the time scale $t_*$ that such paths [Fig.~\ref{fig:Path}(b,c)] begin to dominate. At earlier times, the behavior of the SFF is controlled by paths that do not revisit basis states [Fig.~\ref{fig:Path}(a)].

The way that we have introduced non-unitarity should be contrasted with that in recent studies of many-body quantum systems with weak dissipation \cite{mori2024liouvillian,zhang2025thermalization,jacoby2025spectral,yoshimura2025theory}. In our case, the non-unitary evolution operator does not describe physical quantum dynamics, for example pure states evolve into pure states, but their norms are not preserved. Introducing dissipation, on the other hand, causes normalized pure states to evolve into normalized mixed states. Dissipation has the effect of suppressing contributions to density matrices from operators that have large support, effectively eliminating part of the (doubled) Hilbert space in which the density matrix evolves \cite{prosen2002ruelle,rakovszky2022dissipation}. Our non-unitary fields instead enhance and suppress amplitudes associated with different basis states. They can be viewed as eliminating part of the Hilbert space in which pure states evolve.

We have uncovered an interesting parallel between these two settings in the limit where the degree of non-unitarity goes to zero. When this non-unitarity describes physical dissipation, taking the thermodynamic limit $N \to \infty$ and subsequently sending the dissipation strength to zero does \emph{not} recover unitary evolution (which instead emerges when the dissipation is sent to zero before the thermodynamic limit is taken) \cite{prosen2002ruelle,mori2024liouvillian,zhang2025thermalization,jacoby2025spectral,yoshimura2024robustness}. In this work we have identified a non-commutation of limits in a different class of non-unitary systems. For example, we have shown that computing the SFF at large $N$ and then sending the strength $h$ of the non-unitary field to zero leads to different behavior compared with sending $h \to 0$ followed by $N \to \infty$ (where one finds unitary evolution). The SFF in the analogous limit for dissipative systems was recently evaluated in Refs.~\cite{yoshimura2024robustness,yoshimura2025theory}. Our results therefore indicate that the non-commutation of limits encountered in dissipative systems is just one example of a much more general phenomenon.

As we have discussed, there are also clear connections between our results and studies of purification transitions in monitored quantum dynamics \cite{Gullans2020,Choi2020}. In that context it is known that, when evolution is weakly non-unitary (for example, because measurements are performed at a low rate), a mixed state can take an amount of time that is exponential in system size to purify. In other words, the system remains sensitive to its initial conditions over a long period of time. One motivation for studying this phenomenon in a system with a fixed evolution operator has been to establish a connection between that behavior and spectral statistics. Previous studies have investigated slow purification, and purification transitions, in systems whose time evolution is generated by a fixed non-Hermitian Hamiltonian with PT symmetry~\cite{Sarang2021,jian2021yang}. In a system with PT symmetry, eigenvalues come in complex conjugate pairs, and slow purification can arise because real Hamiltonian eigenvalues must coalesce before developing imaginary components. Because the evolution operators that describe generic monitored dynamics do not have PT symmetry, it is unclear whether there is a connection between the mechanisms encountered in these two settings. Here we studied fixed evolution operators that do not have PT symmetry, and we have shown that slow purification arises from the fact that the distribution of eigenvalue magnitudes has sharp edges. 

We anticipate that our results will be useful for understanding various features of translation-invariant tensor networks, from the computational complexity of their contraction \cite{schuch2007computational,gonzalez2024random,chen2025sign,tang2025matrix} to their spectral properties \cite{shivam2023many}. However, the evolution operators studied here do not have spatial structure, and the time scale $t_*$ relevant to purification is exponential in system size for all finite non-unitary field strengths. For fixed non-unitary evolution operators \emph{with} spatial structure, it is natural to ask if there are entanglement and purification transitions as the non-unitary field strength is increased, analogous to the situation in monitored random circuits \cite{li2018zeno,skinner2019measurement}. The existence of a volume-law entangled phase would suggest the existence of large classes of translation-invariant tensor networks that are exponentially costly (in system linear dimension) to contract using classical computers, even when their transfer matrices are away from the unitary limit. Such a result would have deep implications for the study of ground states of two-dimensional quantum systems. 

\begin{acknowledgments}
The authors are grateful to Alex Jacoby, Sarang Gopalakrishnan, and Adam Nahum for useful discussions. This work was supported by the Gordon and Betty Moore Foundation (SJG), the NSF QLCI program through Grant No. OMA-2016245 (EA), and a Simons Investigator Award (EA). Numerical calculations were carried out using services provided by the OSG Consortium \cite{osg1,osg2,osg3,osg4}, which is supported by NSF awards 2030508 and 2323298. 
\end{acknowledgments}

\appendix

\renewcommand{\theequation}{\thesection\arabic{equation}}

\section{Purification dynamics}
\label{sec:appA}
In this appendix, we derive the early-time logarithmic decay of entropy for maximally mixed initial state in the models $\tilde{T}(t)$ and $T(t)=T^t$, respectively. By identifying the subleading contributions, we define the onset time of the deviation from this early-time behavior, highlighting the importance of time-translation invariance.

\subsection{Without time-translation invariance}
\label{sec:appA1}
To compute the second R\'enyi entropy with the replica trick~[Eq.~(\ref{eq:S2main})], we focus on two kinds of ensemble-averaged moments of singular values $\langle\text{Tr}[(\tilde{T}(t)\tilde{T}^\dagger(t))^{2k}]\rangle$ and $\langle\text{Tr}[(\tilde{T}(t)\tilde{T}^\dagger(t))^k]^2\rangle$, where $\tilde{T}(t)=\prod_{l=1}^{t}\zeta U_l$ with independent Haar-random $U_l$. Before proceeding, it is instructive to look at the general structure of the ensemble-averaged $\langle\text{Tr}[(\tilde{T}(t)\tilde{T}^\dagger(t))^m]^n\rangle$ involving $m$ copies of $\tilde{T}(t)\tilde{T}^\dagger(t)$ within each copy of $n$ traces.

By considering the eigenbasis of $\zeta$ and inserting resolutions of identity into $\langle\text{Tr}[(\tilde{T}(t)\tilde{T}^\dagger(t))^m]^n\rangle$, there are $mnt$ basis states $\{a_{i,j;l},a_{i,j;l}^*\}_{i=1,\dots,m;\ j=1,\dots,n;\ l=1,\dots,t}$ in the `forward' and `backward' directions, corresponding to $\tilde{T}(t)$ and $\tilde{T}^\dagger(t)$, respectively. Specifically, we have
\begin{widetext}
\begin{align}
\text{Tr}[(\tilde{T}(t)\tilde{T}^\dagger(t))^m]^n=\prod_{j=1}^{n}\prod_{i=1}^{m}\sum_{\{a_{i,j;l},a_{i,j;l}^*\}_{l=1}^{t}}\begin{aligned}[t]&\zeta_{a_{i,j;t}}U_{t,a_{i,j;t}a_{i,j;t-1}}\cdots\zeta_{a_{i,j;2}}U_{2,a_{i,j;2}a_{i,j;1}}\zeta_{a_{i,j;1}}\delta_{a_{i,j;1}a_{i,j;1}^*}
\\
&\times\zeta_{a_{i,j;1}^*}U_{2,a_{i,j;2}^*a_{i,j;1}^*}^*\cdots\zeta_{a_{i,j;t-1}^*}U_{t,a_{i,j;t}^*a_{i,j;t-1}^*}^*\delta_{a_{i+1,j;t}a_{i,j;t}^*}.\end{aligned}
\end{align}
\end{widetext}
Due to the structure of the trace, there are $mn$ identity permutations, $a_{i,j;1}=a_{i,j;1}^*$, at $l=1$ time step and $n$ swap permutations over $m$ basis states, $a_{i+1,j;t}=a_{i,j;t}^*$, at $l=t$ time step.

Note also that, since here the unitary operators $U_l$ at different time steps are statistically independent, we only have `equal-time' pairings of indices. Explicitly, basis states at a certain time step in the `forward' direction, $\{a_{i,j;l}\}_{i=1\dots m;\ j=1\dots n}$, must be paired with basis states at identical time step in the `backward' direction, $\{a_{i,j;l}^*\}_{i=1\dots m;\ j=1\dots n}$. Equivalently, the permutations $\sigma,\tau$ over $mnt$ basis states are factorized as $\sigma=\prod_l\sigma_l$~($\tau=\prod_l\tau_l$), where $\sigma_l,\tau_l\in\mathcal{S}_{mn}$. Thus, the ensemble-averaged $\langle\text{Tr}[(\tilde{T}(t)\tilde{T}^\dagger(t))^m]^n\rangle$ is given by
\begin{align}
\langle\text{Tr}&[(\tilde{T}(t)\tilde{T}^\dagger(t))^m]^n\rangle\label{eq:Moment}
\\
=& \prod_{j=1}^{n}\prod_{i=1}^{m}\sum_{\{a_{i,j;l},a_{i,j;l}^*\}_{l=1}^{t}}\delta_{a_{i,j;1}a_{i,j;1}^*}\delta_{a_{i,j;t}a_{i+1,j;t}^*}\nonumber
\\
&\times\prod_{l=1}^{t-1}\sum_{\sigma_l,\tau_l\in \mathcal{S}_{mn}}\begin{aligned}[t]&\text{Wg}(\sigma_l\tau_l^{-1})\zeta_{a_{i,j;l}}\zeta_{a_{i,j;l}^*}
\\
&\times\delta_{a_{i,j;l}a_{\sigma_l({i,j});l}^*}\delta_{a_{i,j;l+1}a_{\tau_l(i,j);l+1}^*},\end{aligned}\nonumber
\end{align}
which depends on the cycle structure of $\sigma_l\tau_l^{-1}$ and the pairing of basis states at the same time step. Defining $\tau_0=(\text{Id}_m)^{\otimes n}$ and $\sigma_t=(\text{SWAP}_m)^{\otimes n}$, where $\text{Id}_m$ and $\text{SWAP}_m$ act on $m$ copies of $\tilde{T}(t)\tilde{T}^\dagger(t)$ within each copy of $n$ traces, the pairing of $\{a_{i,j;l},a_{i,j;l}^*\}$ depends on the cycle structure of $\sigma_l\tau_{l-1}^{-1}$ for $l=1,\dots,t$. Suppose $\sigma_l\tau_{l-1}^{-1}=\{n_{l,k}\}_{k=0}^{m_l-1}$, i.e. the pairing at time $l$ factorizes the $mn$ basis states into $m_l\leq mn$ independent basis states subject to $\sum_{k=0}^{m_l-1}n_{l,k}=mn$, and each independent basis state is visited $n_{l,k}$ times individually in the forward and backward directions. Each such cycle generates a factor $\text{Tr}[\zeta^{2n_{l,k}}]$, and we refer to this as an $n_{l,k}$-loop contribution (for example, a $2$-loop contributes $\text{Tr}[\zeta^4]$). This simplifies Eq.~(\ref{eq:Moment}) as
\begin{align}
\langle\text{Tr}[(\tilde{T}(t)\tilde{T}^\dagger(t))^m]^n\rangle&=\sum_{\sigma_l,\tau_l\in\mathcal{S}_{mn}}\Big(\prod_{k=1}^{m_l}\text{Tr}[\zeta^{2n_{l,k}}]\Big)
\\
&\times\prod_{l=1}^{t-1}\bigg[\text{Wg}(\sigma_l\tau_l^{-1})\prod_{k=1}^{m_l}\text{Tr}[\zeta^{2n_{l,k}}]\bigg],\nonumber
\end{align}
where $n_{l,k}$ and $m_{l}$ are properties of $\sigma_l\tau_{l-1}^{-1}$.
Hence the leading contribution arises from the permutations that maximize both $\text{Wg}(\sigma_l\tau_l^{-1})$ and the number of independent pairings $m_l$.

Here we show that permutations generating more domain walls within each copy of trace only become significant at time scale $t\sim t_*^2$ by calculating $\langle\text{Tr}[(\tilde{T}(t)\tilde{T}^\dagger(t))^m]^n\rangle$ at $m=2$ and $n=1$, where several leading contributions are given by
\begin{widetext}
\begin{align}
\langle\text{Tr}[(\tilde{T}(t)\tilde{T}^\dagger(t))^2]\rangle= & \,t\text{Wg}(1^2)^{t-1}\text{Tr}[\zeta^2]^{2t-2}\text{Tr}[\zeta^4]+(t-1)\text{Wg}(2)\text{Wg}(1^2)^{t-2}\text{Tr}[\zeta^2]^{2t}\nonumber
\\
& +\frac{t^3}{6}\begin{aligned}[t]\bigg[ & \text{Wg}(1^2)^{t-1}\text{Tr}[\zeta^2]^{2t-6}\text{Tr}[\zeta^4]^3+3\text{Wg}(1^2)^{t-2}\text{Wg}(2)\text{Tr}[\zeta^2]^{2t-4}\text{Tr}[\zeta^4]^2
\\
& +3\text{Wg}(1^2)^{t-3}\text{Wg}(2)^2\text{Tr}[\zeta^2]^{2t-2}\text{Tr}[\zeta^4]+\text{Wg}(1^2)^{t-4}\text{Wg}(2)^3\text{Tr}[\zeta^2]^{2t}\bigg]+\mathcal{O}(D^{-3})\end{aligned}\nonumber
\\
= & \,D\mathbb{E}[\zeta^2]^{2t}\bigg[1+t\bigg(\frac{\mathbb{E}[\zeta^4]}{\mathbb{E}[\zeta^2]^2}-1\bigg)+\frac{t^3}{6}D^{-2}\bigg(\frac{\mathbb{E}[\zeta^4]}{\mathbb{E}[\zeta^2]^2}-1\bigg)^3+\mathcal{O}(D^{-4})\bigg]\nonumber
\\
\approx & \,D\mathbb{E}[\zeta^2]^{2t}\bigg[1+t\bigg(\frac{\mathbb{E}[\zeta^4]}{\mathbb{E}[\zeta^2]^2}-1\bigg)\bigg]\text{\ \ at\ \ }t\ll D\Big(\frac{\mathbb{E}[\zeta^4]}{\mathbb{E}[\zeta^2]^2}-1\Big)^{-1}\text{\ \ and\ \ }D\gg1.
\end{align}
\end{widetext}
Since the boundary condition of $\text{Tr}[\tilde{T}(t)\tilde{T}^\dagger(t)\tilde{T}(t)\tilde{T}^\dagger(t)]$ is given by $\text{Id}_2$ and $\text{SWAP}_2$ at $l=1$ and $l=t$~[Fig.~\ref{fig:Random}(a)], the leading diagram~($\sigma=\tau$ having maximal number of disjoint cycles) can only involve $\text{Tr}[\zeta^4]\text{Tr}[\zeta^2]^{2t-2}$~(cycle structure $21^{2t-2}$), where there are $t$ locations to insert the domain wall. This determines the contribution $t\text{Wg}(1^2)^{t-1}\text{Tr}[\zeta^2]^{2t-2}\text{Tr}[\zeta^4]\sim\mathcal{O}(D)$. Next, by considering a single elementary transposition on $2$-loop at the domain wall, we have $t-1$ distinct combinations of $\sigma,\tau$, resulting in $(t-1)\text{Wg}(2)\text{Wg}(1^2)^{t-2}\text{Tr}[\zeta^2]^{2t}$.

The next-to-leading contribution is $\mathcal{O}(D^{-1})$ since the number of domain walls at $\sigma=\tau$ is always odd. In the diagonal approximation~($\sigma=\tau$, $\mathcal{O}(D^{-2(t-1)})$), it corresponds to diagrams with cycle structures $\{2^31^{2t-6}\}$, where the number of disjoint cycles is $2t-3$ of $\mathcal{O}(D^{2t-3})$. This type of diagram can be regarded as three domain walls, or four partitions over $t$ indices. For general number of domain walls $w$, denote the length of each partition as $t_{0,\cdots,w}$ with a constraint $\sum_{j=0}^{w}t_j=t-w$, then the number of configurations of $w$ domain walls is given by $\binom{(w+1)+(t-w)-1}{t-w}=\binom{t}{t-w}\approx t^w/w!$. At $w=3$ we have contribution $(t^3/6)\text{Wg}(1^2)^{t-1}\text{Tr}[\zeta^2]^{2t-6}\text{Tr}[\zeta^4]^3$.

Given a permutation $\sigma$ with cycle structure $\{2^31^{2t-6}\}$, there are three types of $\tau\neq\sigma$ of order $D^{-1}$. The first type is one elementary transposition on three $2$-loops of three choices, leading to cycle structure $\{2^21^{2t-4}\}$ with contribution $(t^3/6)3\text{Wg}(1^2)^{t-2}\text{Wg}(2)\text{Tr}[\zeta^2]^{2t-4}\text{Tr}[\zeta^4]^2$. The second type is two elementary transpositions on three $2$-loops of three choices, leading to cycle structure $\{21^{2t-2}\}$ with contribution $(t^3/6)3\text{Wg}(1^2)^{t-3}\text{Wg}(2)^2\text{Tr}[\zeta^2]^{2t-2}\text{Tr}[\zeta^4]$. The final one is three elementary transpositions on three $2$-loops of unique choice, leading to cycle structure $\{1^{2t}\}$ with contribution $(t^3/6)\text{Wg}(1^2)^{t-4}\text{Wg}(2)^3\text{Tr}[\zeta^2]^{2t}$.

Combining these expressions, we find that contributions at same order of $D$ come in series of $\text{Tr}[\zeta^4]/\text{Tr}[\zeta^2]^2-D^{-1}=t_*^{-2}$, so that they vanish in the unitary limit. This expression arises since performing a single elementary transposition on $2$-loop at the domain wall~(for $\sigma=\tau$) changes the contribution from $\text{Wg}(1^2)\text{Tr}[\zeta^4]$ to $\text{Wg}(1^2)\text{Tr}[\zeta^2]^2$, corresponding to two $1$-loops. In the following we will focus on the leading contribution with $\sigma=\tau$ and replace $\text{Tr}[\zeta^4]/\text{Tr}[\zeta^2]^2$ by $t_*^{-2}$ to include the contributions of $\sigma\neq\tau$ at the same order of $D$. In principle there are $2^{t-1}$ distinct $\tau$ for each $\sigma$, while other $\tau$ would result in higher order suppression from Weingarten function, which is negligible.

As we show in Sec.~\ref{sec:purify}, $\langle\text{Tr}[\tilde{T}(t)\tilde{T}^\dagger(t)]^2\rangle\approx\langle\text{Tr}[\tilde{T}(t)\tilde{T}^\dagger(t)]\rangle^2\approx D^2\mathbb{E}[\zeta^2]^{2t}$ at $t\ll t_*^2$ and $D\gg 1$ implies that contributions with additional domain walls across copies of traces become significant only for $t \sim t_*^2$. This suggests that the leading contributions to $\langle\text{Tr}[(\tilde{T}(t)\tilde{T}^\dagger(t))^2]^k\rangle$ and $\langle\text{Tr}[(\tilde{T}(t)\tilde{T}^\dagger(t))]^{2k}\rangle$ at $t\ll t_*^2$ arise from permutations $\sigma=\tau$ within each copy of trace, which generate the least number of domain walls compatible with the boundary conditions. As a result, these averages are factorized over copies of traces and approximated as $\langle\text{Tr}[(\tilde{T}(t)\tilde{T}^\dagger(t))^2]^k\rangle\approx \langle\text{Tr}[(\tilde{T}(t)\tilde{T}^\dagger(t))^2]\rangle^k$ and $\langle\text{Tr}[(\tilde{T}(t)\tilde{T}^\dagger(t))]^{2k}\rangle\approx\langle\text{Tr}[(\tilde{T}(t)\tilde{T}^\dagger(t))]\rangle^{2k}$.
Hence, the early-time second R\'enyi entropy is given by the replica trick
\begin{widetext}
\begin{align}
\langle \tilde{S}_2(t)\rangle= & \lim_{k\to0}\frac{1}{k}\bigg[\langle\text{Tr}[\tilde{T}(t)\tilde{T}^\dagger(t)]^{2k}\rangle-\langle\text{Tr}[(\tilde{T}(t)\tilde{T}^\dagger(t))^2]^k\rangle\bigg]\nonumber
\\
\approx & \lim_{k\to 0}\frac{1}{k}\bigg[\bigg(D^2\mathbb{E}[\zeta^2]^{2t}\bigg)^k-\bigg(D\mathbb{E}[\zeta^2]^{2t}\Big[1+t\bigg(\frac{\mathbb{E}[\zeta^4]}{\mathbb{E}[\zeta^2]^2}-1\bigg)\Big]\bigg)^k\bigg]=\ln D-\ln\bigg[1+t\bigg(\frac{\mathbb{E}[\zeta^4]}{\mathbb{E}[\zeta^2]^2}-1\bigg)\bigg]
\end{align}
\end{widetext}
at $t\ll t_*^2$ and $D\gg1$. This means that, when there is no time-translation invariance, the early-time entropy decays logarithmically.

\subsection{With time-translation invariance}
\label{sec:appA2}
Now we restore the time-translation invariance, which allows non-equal-time pairing, such that there are $(kt)!$ distinct pairings when there are $k$ copies of $T^t(T^\dagger)^t$. Even though the number of pairings is large, we can focus on the leading contribution from $\sigma=\tau$ with maximal degeneracy~(polynomial in $t$) in series of $D$.
It is evident that the permutations whose cycle structures only involve 2-loop and 1-loop, which respectively correspond to pairings of four and two $\zeta$, have largest degeneracy due to the largest number of independent pairings at the same order of $D$. This is similar to the calculation of SFF, where $\{2^21^{t-4}\}$ appears at shorter time scale than $\{31^{t-3}\}$. Other $\tau\neq \sigma$ in the same order would give rise to expansion in series of $\text{Tr}[\zeta^4]/\text{Tr}[\zeta^2]^2-D^{-1}=t_*^{-2}$.

We begin with the calculation of $\langle\text{Tr}[(T^t(T^\dagger)^t)^2]\rangle$. Note that the time-translation invariance allows pairings of $T$ and $T^\dagger$ at different time steps. The leading contribution is still given by the equal-time pairing as the case without time-translation invariance. However, the next-to-leading contribution of order $\mathcal{O}(D^{-1})$, given by $\sigma=\tau$ associated with three factors of $\text{Tr}[\zeta^4]$, reflects the impact of time-translation invariance. This contribution corresponds to the diagram in Fig.~\ref{fig:TimeTI}(a). The four lengths of partitions with equal-time pairings, $t_{1,2,3,4}$, at the corners fix the locations of two factors of $\text{Tr}[\zeta^4]$. The degree of freedom to place the rest one factor of $\text{Tr}[\zeta^4]$ is now determined by the edge with the least free indices. In the particular case shown in Fig.~\ref{fig:TimeTI}(a), it is the edge with $t_{1,2}$ equal-time pairings on two sides, leading to degeneracy $t-t_1-t_2$. This also sets the upper bound on $t_{3,4}$ as $t_3\leq t_1$ and $t_4\leq t_2$, otherwise other edge would instead decide the degree of freedom to place the last factor of $\text{Tr}[\zeta^4]$. To sum up, the degeneracy of this diagram is~(prefactor $4$ from different choices of edges with maximal $t_a+t_b$)
\begin{align}
&4\sum_{t_{1,2};\text{\ }t_1+t_2\leq t}\sum_{t_3\leq t_2}\sum_{t_4\leq t_1}t-t_1-t_2\nonumber
\\
&\approx 4\int_{t_1+t_2\leq t}dt_1dt_2t_1t_2(t-t_1-t_2)=\frac{t^{5}}{30},\nonumber
\end{align}
which yields
\begin{align}
\langle\text{Tr}[(T^t(T^\dagger)^t)^2]\rangle=& D\mathbb{E}[\zeta^2]^{2t}\bigg[1+t\bigg(\frac{\mathbb{E}[\zeta^4]}{\mathbb{E}[\zeta^2]^2}-1\bigg)
\\
&+\frac{t^5}{30}D^{-2}\bigg(\frac{\mathbb{E}[\zeta^4]}{\mathbb{E}[\zeta^2]^2}-1\bigg)^3+\cdots\bigg]\nonumber
\end{align}
at $t\ll t_*$ and $D\gg1$.

Next, we carry out the similar calculation for $\langle\text{Tr}[T^t(T^\dagger)^t]\rangle$. The leading contribution is still given by the equal-time pairing without domain wall as the case without time-translation invariance. The diagram in Fig.~\ref{fig:TimeTI}(b) corresponds to the next-to-leading contribution from $\sigma=\tau$ with two factors of $\text{Tr}[\zeta^4]$. We denote the number of non-equal-time pairing as $t_1$, which divides these $t_1$ indices into three partitions~(of $\binom{t_1+3-1}{t_1}\approx t_1^2/2$) and leads to $t-t_1$ degeneracy~(free to translate forward or backward between boundaries). This gives the degeneracy $\sum_{t_1\leq t}(t-t_1)t_1^2/2\approx t^4/24$, such that we have
\begin{align}
\langle\text{Tr}[T^t(T^\dagger)^t]\rangle =&\, D\mathbb{E}[\zeta^2]^{t}
\\
&\times\bigg[1+\frac{t^4}{24}D^{-2}\bigg(\frac{\mathbb{E}[\zeta^4]}{\mathbb{E}[\zeta^2]^2}-1\bigg)^2
+\cdots\bigg]\nonumber
\end{align}
at $t\ll t_*$ and $D\gg1$.

\begin{figure}[t!]
\centering{}
\includegraphics[width=0.48\textwidth]{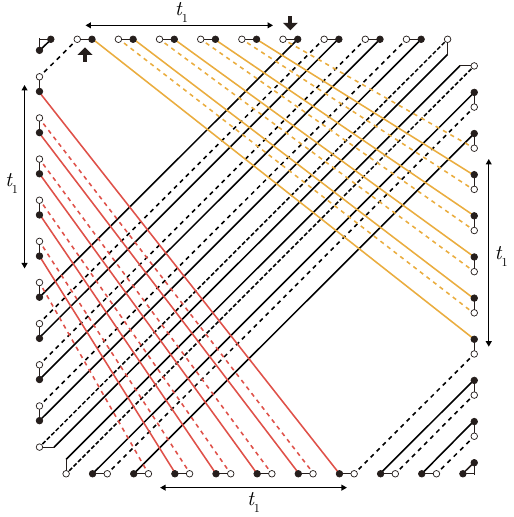}
\caption{\label{fig:TimeTI_factorized}
Schematics of next-to-leading $\sigma=\tau$ involving non-equal-time pairings in the calculation of $\langle\text{Tr}[T^t(T^\dagger)^t]^2\rangle$. The equal-time pairings~(black) are shown in diagonal, and two factors of $\text{Tr}[\zeta^4]$ are indicated by thick arrows. Once the non-equal-time pairings of $T$ and $T^\dagger$ belonging to different copies of traces, such as yellow lines, are determined, the red non-equal-time pairings are uniquely determined since they are embedded within the same sets of equal-time pairings.
}
\end{figure}

Moreover, we can show that permutations across copies of traces only contribute significantly at $t\sim t_*$. We first note that the leading contributions to $\langle\text{Tr}[(T^t(T^\dagger)^t)^2]\rangle$ and $\langle\text{Tr}[T^t(T^\dagger)^t]\rangle$ are given by the equal-time pairing within each copy of trace with one and zero domain walls for $t\ll t_*$. With sparsely-distributed domain walls within each copy of trace, the additional non-equal-time pairings across copies of traces should be embedded within these equal-time pairings. Thus it is enough to consider the next-to-leading contribution from non-equal-time pairings in $\langle\text{Tr}[T^t(T^\dagger)^t]^2\rangle$, which corresponds to the diagram in Fig.~\ref{fig:TimeTI_factorized}. This diagram can be viewed as the unfolded version of Fig.~\ref{fig:Random}(d). We denote the number of non-equal-time-pairings as $t_1$, and these $t_1$ pairings are free to translate forward or backward between boundaries, independently in $T^t$ and $(T^\dagger)^t$. This gives the degeneracy $\sum_{t_1\leq t}(t-t_1)^2\approx t^3/3$, such that we have
\begin{align}
\langle\text{Tr}&[T^t(T^\dagger)^t]^2\rangle =
\langle\text{Tr}[T^t(T^\dagger)^t]\rangle^2
\\
&+D^2\mathbb{E}[\zeta^2]^{2t}\bigg[\frac{t^3}{3}D^{-2}\bigg(\frac{\mathbb{E}[\zeta^4]}{\mathbb{E}[\zeta^2]^2}-1\bigg)^2
+\cdots\bigg]\nonumber
\end{align}
at $t\ll t_*$. It is evident that permutations across different copies of traces are negligible at $t\ll t_*$. Hence the averages are factorized over copies of traces, and the early-time evolution of second R\'enyi entropy is the same as the case without time-translation invariance
\begin{align}
\langle S_2(t)\rangle\approx & \,\ln D-\ln\bigg[1+t\bigg(\frac{\mathbb{E}[\zeta^4]}{\mathbb{E}[\zeta^2]^2}-1\bigg)\bigg]\label{eq:Renyi2}
\end{align}
only at a parametrically shorter time scale $t\ll t_*$ and $D\gg1$.

Our result suggests that time-translation invariance starts to affect the singular value dynamics after the exponentially long time $t_*$. An immediate conclusion is that the bulk distribution of early-time singular values is insensitive to the time-translation invariance.

\subsection{Late-time purification dynamics}
\label{sec:appA3}
In the absence of time-translation invariance and at $t\gg t_*^2$, the entropy is much smaller than unity. Although a diagrammatic calculation of the entropy is complicated, it is possible to focus on the leading two singular values to analyze the late-time purification since singular values at late times are well separated as $\sigma_\alpha(t)\approx tl_\alpha$. Here $l_\alpha$ is a Lyapunov exponent. This behavior is similar to the dynamics involving forced (but random) measurements discussed in Ref.~\cite{nahum2021}, and now we apply it to a deterministic non-unitary setting. Denoting the singular value decomposition as $\tilde{T}(t)=V_t\Sigma(t)W_t^\dagger$, we can compute the leading two singular values at $t+1$ by considering
\begin{align}
\tilde{T}^\dagger(t+1)\tilde{T}(t+1)=W_t\Sigma(t) V_{t}^\dagger U_{t+1}^\dagger\zeta^2U_{t+1}V_t\Sigma(t)W_t^\dagger.
\end{align}
Due to the statistical independence of $U$ at each time step, we define $\tilde{U}=U_{t+1}V_t$ that inherits the Haar randomness of $U_{t+1}$, the squared singular values of $\tilde{T}(t+1)$ are thus given by the eigenvalues of $X(t)=\Sigma(t)\tilde{U}^\dagger\zeta^2\tilde{U}\Sigma(t)$, which is approximated by a $2\times2$ matrix at late times. Although the entries of a Haar random matrix are correlated, in the following these correlations only lead to corrections that are suppressed by powers of $D$ relative to the leading behavior, so we can safely neglect them, i.e. we regard the elements of $X(t)$ as independent Gaussian random variables. Performing a Haar average to obtain the mean and the strength of fluctuations of these matrix elements, we have
\begin{align}
X(t)=\mathbb{E}[\zeta^2]\begin{pmatrix}
e^{2\sigma_0(t)}(1+t_*^{-1} a) & e^{\sigma_0(t)+\sigma_1(t)}t_*^{-1}\beta
\\
e^{\sigma_0(t)+\sigma_1(t)}t_*^{-1}\beta^* & e^{2\sigma_1(t)}(1+t_*^{-1} b)
\end{pmatrix},
\end{align}
where $a$, $b$, $\beta$ have zero mean and $\langle a^2\rangle=\langle b^2\rangle=\langle |\beta|^2\rangle=1$. This results in a recursion relation for the squared ratio of two leading singular values~(which gives the limiting singular value gap as the gap of leading Lyapunov exponents, $\Delta_\sigma\equiv t^{-1}[\sigma_0(t)-\sigma_1(t)]=l_0-l_1$)
\begin{align}
e^{-2(t+1)\Delta_\sigma}\approx e^{-2t\Delta_\sigma}\bigg[\frac{1+t_*^{-1}b}{1+t_*^{-1}a}-\frac{t_*^{-2}|\beta|^2}{(1+t_*^{-1}a)^2}\bigg],\label{eq:Sratio}
\end{align}
where we assume well-separated singular values $t\Delta_\sigma\gg1$ at late times. Since $t_*^{-1}$ is exponentially small in $N$, we can further simplify Eq.~(\ref{eq:Sratio}) as
\begin{align}
\Delta_\sigma\approx \frac{a-b}{2}t_*^{-1}+\Big(\frac{|\beta|^2}{2}-\frac{a-b}{4}\Big)t_*^{-2},\label{eq:Sgap}
\end{align}
whose average determines the limiting singular value gap $\Delta_\sigma=t_*^{-2}/2$. Therefore, for $t \gtrsim t_*^2$, the von Neumann entropy and the $n^{\text{th}}$ R\'enyi entropy are~($n>1$)
\begin{align}
S(t)\approx\frac{t}{t_*^2}e^{-t/t_*^2},\ \ S_n(t)\approx\frac{n}{n-1}e^{-t/t_*^2}.
\end{align}
The time scale $t_*^2$ in this system without time-translation invariance should be contrasted with the purification time $t_P \sim t_*$ found in systems \emph{with} time-translation invariance.

We note that when we restore the time-translation invariance, $V_t$ and $U_{t+1}$ are not statistically independent since the Haar random unitary is identical at each time step, such that the above analysis fails to describe this case. We can nevertheless analyze the late-time purification by Yamamoto's theorem, $\sigma_\alpha(t)\approx t\rho_\alpha$ at long time, as discussed in the main text.

\section{Non-unitary Dyson Brownian motion}
\label{sec:appB}
The spectrum of $T=\zeta U$ can be regarded as one instance of a random matrix arising from a generalized Dyson Brownian motion \cite{dyson1962brownian}. As stated in the main text, this motion can be defined by the update $T(s+ds)=T(s)e^{i Hds}$ with a Gaussian unitary ensemble $H$, satisfying $\overline{H_{ij}}=0$ and $\overline{H_{ij}(s) H_{kl}^*(s')}= \delta_{ik}\delta_{jl}\delta(s-s')$, such that $\prod e^{i Hds}$ resembles a Haar-random unitary $U$. In each time step, the change in eigenvalues $\lambda_m$ of $T(s)$ within the second-order perturbation theory is
\begin{align}
d\lambda_m=&\,i\lambda_m\langle l_m|Hds|r_m\rangle-\frac{1}{2}\lambda_m\langle l_m|H^2ds^2|r_m\rangle\nonumber
\\
&-\sum_{n\neq m}\frac{\lambda_m\lambda_n}{\lambda_m-\lambda_n}\langle l_m|Hds|r_n\rangle\langle l_n|Hds|r_m\rangle.
\end{align}
Averaging over $H$ up to $\mathcal{O}(ds)$ yields
\begin{align}
\overline{d\lambda_m} & =-\frac{2^N}{2} ds\lambda_m-ds\sum_{n\neq m}\frac{\lambda_m\lambda_n}{\lambda_m-\lambda_n},\label{eq:dlambda}
\\
\overline{d\lambda_m d\lambda_n} & =-ds\lambda_m^2\delta_{mn}\text{,\ \ }\overline{d \lambda_m d\lambda_n^*} =ds\lambda_m\lambda_n^*O_{mn},
\end{align}
where $O_{mn}=\langle l_m|l_n\rangle\langle r_n|r_m\rangle$ is the overlap matrix quantifying the non-orthogonality between left and right eigenstates $\langle l_m|$ and $|r_m\rangle$. It is challenging to directly solve these coupled Brownian motions of complex eigenvalues since the evolution of the overlap matrix is extremely complicated
\begin{align}
\overline{d O_{mn}}=ds&\sum_{p\neq m}\sum_{q\neq n}O_{mn}O_{pq}\Bigg[\frac{\lambda_m}{\lambda_m-\lambda_p}\Big(\frac{\lambda_n}{\lambda_n-\lambda_q}\Big)^*\nonumber
\\
&+\frac{\lambda_p}{\lambda_m-\lambda_p}\Big(\frac{\lambda_q}{\lambda_n-\lambda_q}\Big)^*\Bigg].
\end{align}
However, it is possible to extract the essential mechanism of radial level attraction between eigenvalues. Parameterizing the eigenvalues as $\lambda_m=e^{\rho_m+i\phi_m}$, Eq.~(\ref{eq:dlambda}) becomes
\begin{align}
\overline{d \rho_m}&+i\overline{d\phi_m}\nonumber
\\
& = -ds\Big(\frac{\partial}{\partial \rho_m}-i\frac{\partial}{\partial \phi_m}\Big)\begin{aligned}[t]\sum_{n\neq m}\ln\Big(&\cosh(\rho_m-\rho_n)
\\
&-\cos(\phi_m-\phi_n)\Big)\end{aligned}\nonumber
\\
& =-ds\sum_{n\neq m}\frac{\sinh(\rho_m-\rho_n)-i\sin(\phi_m-\phi_n)}{\cosh(\rho_m-\rho_n)-\cos(\phi_m-\phi_n)}.
\end{align}
Since the denominator is always positive, the sign of this drift force is determined by the numerator.
It is evident that the radial and azimuthal forces have opposite signs, reflecting the radial attraction and azimuthal repulsion.

\section{Mean radial eigenvalue density}
\label{sec:appC}
The eigenvalue distribution of the transfer matrix $T$ of dimension $D=2^N$ that we study is invariant under arbitrary biunitary transformations, $T\to VTW^\dagger$ with any pair of unitary matrices $V,W$. For such matrices, it can be shown that eigenspectrum of $T$ forms a dense ring with sharp edges in the complex plane in the $D=2^N\gg1$ regime.

First note that, since the biunitary transformation (the matrices $W$ and $V$) can be chosen so that $VTW^{\dag}$ is the diagonal matrix of singular values of $T$, the eigenvalues of $\zeta U$ are completely determined by the singular values (which are the diagonal entries of $\zeta$ in our case, since $\zeta$ is real). In particular, we will show that the mean radial density of $\zeta U$ is given by the moment-generating function of $\zeta^2$.

\subsection{Large $D$ asymptotic formula}
\label{sec:appC1}
For a Haar-random $U$ and finite Hilbert space dimension $D$, there exists the exact formula of the mean radial probability density based on the supersymmetry method~\cite{Wei2008}. Using the saddle-point method~\cite{Bogomolny2010}, a large $D\gg1$ asymptotic formula is given by
\begin{align}
& D^{-1}n(|\lambda|)=\frac{2}{|\lambda|}v_s(v_s-1)\Theta(v_s)\Theta(1-v_s)\label{eq:LargeL}
\\
& +\frac{1}{|\lambda||\Phi^{\prime\prime}|}\bigg(\text{Erf}\Big[\sqrt{\frac{D|\Phi^{\prime\prime}|}{2}}(1-v_s)\Big]+\text{Erf}\Big[\sqrt{\frac{D|\Phi^{\prime\prime}|}{2}}v_s\Big]\bigg),\nonumber
\end{align}
where $v_s(|\lambda|)$ is the saddle point of $\Phi(v)$ and satisfies
\begin{align}
\Phi(v) & =\ln v+\frac{1}{D}\sum_{j=1}^{D}\ln\Big(1-\frac{v-1}{v|\lambda|^2}\zeta_j^2\Big),
\\
\Phi^{\prime}(v_s) & =\frac{1}{v_s}\bigg[1+\frac{1}{v_s-1}G_{\zeta^2}\Big(\frac{v_s-1}{v_s|\lambda|^2}\Big)\bigg]=0,\label{eq:PhiP}
\\
\Phi^{\prime\prime}(v_s) & =\frac{1}{v_s(v_s-1)}\bigg[1-\frac{1}{v_s^2|\lambda|^2}G_{\zeta^2}^{\prime}\Big(\frac{v_s-1}{v_s|\lambda|^2}\Big)\bigg],\label{eq:PhiPP}
\end{align}
so that $v_s(|\lambda|)$ depends on $\zeta$ and $n(|\lambda|)$ is completely determined by the moment-generating function of $\zeta^2$, $G_{\zeta^2}(u)=\sum_{k=1}^{\infty}u^k\mathbb{E}[\zeta^{2k}]=\frac{1}{D}\sum_{j=1}^D\frac{u\zeta_j^2}{1-u\zeta_j^2}$.

\subsection{Dense ring with sharp edges}
\label{sec:appC2}
Here we focus on the bulk part of the mean radial density. When $0<v_s<1$, both error functions approach $1$ at $D\gg 1$, leading to
\begin{align}
D^{-1}n(|\lambda|)=\frac{2}{|\lambda|}\bigg(\frac{1}{|\Phi^{\prime\prime}(v_s)|}+v_s(v_s-1)\bigg)=\frac{dv_s}{d|\lambda|},\label{eq:Bulk}
\end{align}
where the second equality follows from Eqs.~(\ref{eq:PhiP}) and~(\ref{eq:PhiPP}). Equation~(\ref{eq:Bulk}) suggests that we can regard the saddle point $v_s(|\lambda|)$ as the radial cumulative probability $F(|\lambda|)=\int_0^{|\lambda|}dsn(s)/D$ at $D\gg1$. That is, $v_s=0$ and $v_s=1$ define the inner and outer bulk edges of density of eigenvalues in the complex plane. Specifically, the outer bulk edge can be obtained from the parametric form of Eq.~(\ref{eq:PhiP}), i.e. defining $\epsilon=\frac{1-v_s}{v_s|\lambda|^2}\ll1$ in the vicinity of outer edge and performing expansion in series of $\epsilon$
\begin{align}
& v_s=1+G_{\zeta^2}(-\epsilon)= 1-\epsilon\mathbb{E}[\zeta^2]+\epsilon^2\mathbb{E}[\zeta^4]+\mathcal{O}(\epsilon^3),\label{eq:Para1}
\\
& |\lambda|^2=\frac{1-v_s}{\epsilon v_s}= \mathbb{E}[\zeta^2]-\epsilon\Big(\mathbb{E}[\zeta^4]-\mathbb{E}[\zeta^2]^2\Big)+\mathcal{O}(\epsilon^2),\label{eq:Para2}
\\
& D^{-1}n(\ln|\lambda|)\begin{aligned}[t]&=2|\lambda|^2\frac{dv_s}{d\epsilon}\bigg(\frac{d|\lambda|^2}{d\epsilon}\bigg)^{-1}
\\
&= \frac{2\mathbb{E}[\zeta^2]^2}{\mathbb{E}[\zeta^4]-\mathbb{E}[\zeta^2]^2}+\mathcal{O}(\epsilon).\end{aligned}\label{eq:Para3}
\end{align}
The outer edge $\rho_N=\frac{1}{2}\ln\mathbb{E}[\zeta^2]$ follows from Eq.~(\ref{eq:Para2}), where the mean radial density is $n(\rho_N)$ in the $D\to\infty$ limit. Similarly, the inner bulk edge $-\rho_N=-\frac{1}{2}\ln\mathbb{E}[\zeta^{-2}]$ can be obtained by considering $v_s$ close to zero with $t=\frac{1-v_s}{v_s|\lambda|^2}\gg1$.

\subsection{Exponential suppression of tail}
\label{sec:appC3}
The above behavior of bulk distribution is obtained by taking the $D\to\infty$ limit followed by $\rho\to\rho_N^{-}$. If we consider the finite but large $D$ regime and approach the bulk edges from outside, we can construct the large $D$ asymptotic expression of tail in the regime $\rho>\rho_N$. In this case, the error function with $v_s$ still approaches $1$, while the other error function with $1-v_s$ varies drastically in the vicinity of $\rho_N$. To capture this, we consider again the parametric expansion to have
\begin{align}
&1-v_s\approx \epsilon\mathbb{E}[\zeta^2]\approx\frac{\mathbb{E}[\zeta^2]^2}{\mathbb{E}[\zeta^4]-\mathbb{E}[\zeta^2]^2}\Big(\ln\mathbb{E}[\zeta^2]-\ln|\lambda|^2\Big)<0,\label{eq:Deviation}
\\
& |\Phi^{\prime\prime}|\approx\Big|v_s(1-v_s)+\frac{|\lambda|}{2}\frac{dv_s}{d|\lambda|}\Big|^{-1}\approx \frac{\mathbb{E}[\zeta^4]-\mathbb{E}[\zeta^2]^2}{\mathbb{E}[\zeta^2]^2}
\end{align}
where the approximate equality indicates that we restrict to leading order in $t$. As Eq.~(\ref{eq:Deviation}) captures the most of $\rho=\ln|\lambda|$ dependence around the outer bulk edge $\rho_N=\frac{1}{2}\ln\mathbb{E}[\zeta^2]$, Eq.~(\ref{eq:LargeL}) around $\rho_N$ is approximated by
\begin{align}
n(\rho)\approx & \,\frac{D\mathbb{E}[\zeta^2]^2}{\mathbb{E}[\zeta^4]-\mathbb{E}[\zeta^2]^2}
\\
&\times\bigg(1-\text{Erf}\bigg[\sqrt{\frac{2D\mathbb{E}[\zeta^2]^2}{\mathbb{E}[\zeta^4]-\mathbb{E}[\zeta^2]^2}}\Big(\rho-\frac{1}{2}\ln\mathbb{E}[\zeta^2]\Big)\bigg]\bigg).\nonumber
\end{align}
This tail is suppressed exponentially in $N$, reflecting that density of eigenvalues in the complex plane forms dense ring with sharp edges. It is evident that the mean radial density at the outer bulk edge $n(\rho_N)=D\mathbb{E}[\zeta^2]^2/(\mathbb{E}[\zeta^4]-\mathbb{E}[\zeta^2]^2)$ becomes half of Eq.~(\ref{eq:Para3}). This can be understood from the ordering of limits, i.e. $\lim_{\rho\to\rho_N}\lim_{N\to\infty}n(\rho)=2^{N+1}\mathbb{E}[\zeta^2]^2/(\mathbb{E}[\zeta^4]-\mathbb{E}[\zeta^2]^2)$ recovers Eq.~(\ref{eq:Para3}). Nevertheless, at finite $N$ the mean radial density at the outer bulk edge is still given by
\begin{align}
n(\rho_N)=\frac{2^N\mathbb{E}[\zeta^2]^2}{\mathbb{E}[\zeta^4]-\mathbb{E}[\zeta^2]^2}=t_*^2.
\end{align}

\subsection{Distribution of leading eigenvalue gap}
\label{sec:appC4}
To obtain the distribution of leading eigenvalue gap $\Delta_\lambda$, we focus on $\mathcal{N}=\sqrt{n(\rho_N)/(2\pi)}$ eigenvalues at $\rho\geq\rho_N$ with a normalized probability density
\begin{align}
\tilde{f}(\rho)&=\mathcal{N}^{-1}n(\rho)
\\
&=\sqrt{2\pi n(\rho_N)}\bigg(1-\text{Erf}\Big[\sqrt{2n(\rho_N)}(\rho-\rho_N)\Big]\bigg),\nonumber
\end{align}
such that $\int_{\rho_N}^{\infty}d\rho\tilde{f}(\rho)=1$. Suppose these $\mathcal{N}$ eigenvalue moduli are independently and identically distributed from $\tilde{f}(\rho)$ as we only consider the radial part, then the distribution of leading eigenvalue gap $\Delta_\rho$ is
\begin{align}
f(\Delta_\rho) & =\mathcal{N}(\mathcal{N}-1)\int_{\rho_N}^{\infty}d\rho\tilde{f}(\rho)\tilde{f}(\rho+\Delta_\rho)\big[\tilde{F}(\rho)\big]^{\mathcal{N}-2},\label{eq:LeadingGap}
\end{align}
where $\tilde{F}(\rho)=\int_{\rho_N}^{\rho}ds\tilde{f}(s)$ is the cumulative distribution of $\tilde{f}(\rho)$. Since $\mathcal{N}$ is exponentially large in $N$, $f(\Delta_\rho)$ is thus controlled by the behavior of $\tilde{F}(\rho)$ at $\rho\gg \rho_N$
\begin{align}
\big[\tilde{F}(x)\big]^{\mathcal{N}} & =\Big[1-e^{-x^2}+\sqrt{\pi}x(1-\text{Erf}(x))\Big]^{\mathcal{N}}\nonumber
\\
& \approx \text{exp}\Big[-\frac{\mathcal{N}}{2x^2}e^{-x^2}\Big],\nonumber
\end{align}
where $x=\sqrt{2n(\rho_L)}(\rho-\rho_N)$ and $\tilde{\Delta}_\rho=\sqrt{2n(\rho_N)}\Delta_{\rho}$. By performing the saddle-point method, we can rewrite Eq.~(\ref{eq:LeadingGap}) as
\begin{align}
f(\tilde{\Delta}_\rho) & \approx \mathcal{N}^2\int_0^\infty dx\,\begin{aligned}[t]\text{exp}\bigg[&-\bigg(x^2+(x+\tilde{\Delta}_\rho)^2+\ln x
\\
&+\ln(x+\tilde{\Delta}_\rho)+\frac{\mathcal{N}}{2x^2}e^{-x^2}\bigg)\bigg]\end{aligned}\nonumber
\\
& \approx\begin{aligned}[t]\mathcal{N}^2\int_0^\infty dx\,\text{exp}\bigg[&-\bigg(4x_0^2\Big(x-x_0\Big)^2+2x_0^2+2
\\
+2\ln x_0&+\Big(2x_0+\frac{1}{x_0}\Big)\tilde{\Delta}_\rho+\tilde{\Delta}_\rho^2\bigg)\bigg],\end{aligned}\nonumber
\end{align}
where $x_0=\sqrt{W_0(\mathcal{N}/4)}$ is the solution to the saddle-point equation and $W_0$ denotes the principal branch of the Lambert $W$ function. This leads to an approximately exponential distribution 
\begin{align}
f(\tilde{\Delta}_\rho)&\propto \text{exp}\Big[-\big(2x_0+x_0^{-1}\big)\tilde{\Delta}_\rho\Big]
\\
&\approx\text{exp}\Big[-2\sqrt{\ln\mathcal{N}}\tilde{\Delta}_\rho\Big]\nonumber
\end{align}
and the associated averaged eigenvalue gap
\begin{align}
\langle\Delta_\rho\rangle^{-1}\sim \sqrt{n(\rho_N)\ln\big[n(\rho_N)\big]}.
\end{align}
We note that the saddle-point approximation only works for extremely large $N$. For intermediate values of $N$ accessible in our numerics, we fit the prefactor and find that $\langle\Delta_\rho\rangle^{-1}=B\sqrt{ n(\rho_N)\ln\big[n(\rho_N)\big]}$ with $B \approx 1.77$ provides a better prediction.

\begin{figure*}[t!]
\centering{}
\includegraphics[width=\textwidth]{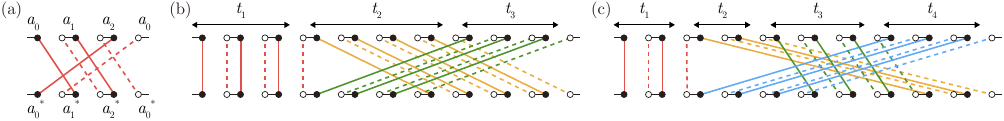}
\caption{\label{fig:Cycle}
(a)~Schematics of $\sigma=\tau$ being cyclic permutations among $t=3$ basis states. The solid and dashed lines represent the permutations $\sigma$ and $\tau$ among $\{a_j\}$ and $\{a_j^*\}$, respectively. In this particular example, we have $a_0=a_1^*$, $a_1=a_2^*$, and $a_2=a_0^*$, leading to three disjoint $1$-loops or $\sigma \mathcal{T}^{-1}\sigma^{-1}\mathcal{T}=\{1^3\}$.
(b)~Schematics of $\sigma \mathcal{T}^{-1}\sigma^{-1}\mathcal{T}=\{31^{t-3}\}$, where $\{a_j\}$ is divided into three partitions $\{t_1t_2t_3\}$ and $\{a_j^*\}$ becomes $\{t_1t_3t_2\}$.
(c)~Schematics of $\sigma \mathcal{T}^{-1}\sigma^{-1}\mathcal{T}=\{2^21^{t-4}\}$, where $\{a_j\}$ and $\{a_j^*\}$ are divided into $\{t_1t_2t_3t_4\}$ and $\{t_1t_4t_3t_2\}$, respectively.
}
\end{figure*}

\section{Spectral form factor}
\label{sec:appD}
In this appendix, we provide details of the calculation of the SFF in the main text. After taking the average over an ensemble of Haar random $U$, the ensemble-averaged SFF in Eq.~(\ref{eq:avgSFF}) is given by the summation over all permutations $\sigma,\tau\in \mathcal{S}_{t}$. In the main text we further simplify Eq.~(\ref{eq:avgSFF}) by considering permutations $\sigma,\tau$ that factorize $2t$ basis states $\{a_k,a_k^*\}_{k=0}^{t-1}$ into $m\leq t$ independent basis states $\{n_k\}_{k=0}^{m-1}$. In fact, this corresponds to the cycle structure of $\sigma\mathcal{T}^{-1}\tau^{-1}\mathcal{T}$, where $\mathcal{T}$ is the translation operator and acts as $a_{\mathcal{T}(r)}=a_{r+1}$. Here $\sigma\mathcal{T}^{-1}\tau^{-1}\mathcal{T}=\{n_k\}_{k=0}^{m-1}$ represents that $\sigma\mathcal{T}^{-1}\tau^{-1}\mathcal{T}$ consists of $m$ disjoint cycles of size $n_k$. Therefore, the ensemble-averaged SFF depends on the cycle structures of $\sigma\tau^{-1}$ and $\sigma\mathcal{T}^{-1}\tau^{-1}\mathcal{T}$.

Here we will employ the diagrammatic expansion based on pairs of permutations to identify the subleading contributions. To this end, it is instructive to review the leading contribution from permutations satisfying $\sigma=\tau$ and $\sigma\mathcal{T}^{-1}\tau^{-1}\mathcal{T}=\{1^t\}$, i.e. $\sigma=\tau$ being cyclic permutations over $t$ basis states. The corresponding diagram is shown in Fig.~\ref{fig:Cycle}.

\subsection{Diagonal contribution}
In the unitary limit, both the deviations from $\sigma\tau^{-1}=\{1^t\}$ and $\sigma \mathcal{T}^{-1}\tau^{-1}\mathcal{T}=\{1^t\}$ results in suppression in power of $D$. In contrast, at $h>0$, the suppression in deviations from $\sigma \mathcal{T}^{-1}\tau^{-1}\mathcal{T}=\{1^t\}$ decays more slowly than the deviations from from $\sigma\tau^{-1}=\{1^t\}$. This allows us to focus on the diagonal contribution with identical $\sigma=\tau$ in the SFF under the approximation $\text{Wg}(1^t)\approx D^{-t}$
\begin{align}
\langle K(t)\rangle\Big|_{\sigma=\tau} & \approx D^{-t}\sum_{\sigma \mathcal{T}^{-1}\sigma^{-1}\mathcal{T}}\prod_{k=0}^{m-1}\text{Tr}[\zeta^{2n_k}]
\\
& =\mathbb{E}[\zeta^2]^t\begin{aligned}[t]\bigg[&\sum_{\{1^t\}}1+D^{-2}\sum_{\{31^{t-3}\}}\frac{\displaystyle\mathbb{E}[\zeta^6]}{\displaystyle \mathbb{E}[\zeta^2]^3}
\\
&+D^{-2}\sum_{\{2^21^{t-4}\}}\frac{\displaystyle \mathbb{E}[\zeta^4]^2}{\displaystyle \mathbb{E}[\zeta^2]^4}+\cdots\bigg],\end{aligned}\nonumber
\end{align}
where the expansion is based on the cycle structure of $\sigma \mathcal{T}^{-1}\sigma^{-1}\mathcal{T}$. Naively we might think that $\{31^{t-3}\}$ is the next-to-leading term and $\{2^21^{t-4}\}$ is the next-to-next-to-leading one since $\mathbb{E}[\zeta^6]\mathbb{E}[\zeta^2]\leq\mathbb{E}[\zeta^4]^2$, while the time scales for them to appear also depend on the number of permutations with the corresponding cycle structures. As we will show in the following, the $\{2^21^{t-4}\}$ term becomes the next-to-leading contribution to SFF.

The number of each permutation with certain cycle structure is constructed as follow:

{\color{darkGreen} $\{31^{t-3}\}$.---}
Suppose we divide two copies of $t$ basis states $\{a_j\}$, $\{a_j^*\}$ into three partitions of length $t_j$ as $\{a_j\}=\{t_1t_2t_3\}$ and $\{a_j^*\}=\{t_1t_3t_2\}$ with a constraint $\sum_{j=1}^3t_j=t$~[Fig.~\ref{fig:Cycle}(b)]. The permutations $\sigma$ that connect partitions of the same length give rise to the $\{31^{t-3}\}$ cycle structure. To avoid overcounting, we define $t_1$ as the length of partition involving $a_0$.

The number of this type of permutations is given by $t\sum_{t_1\leq t}t_1\binom{t-t_1+2-1}{t-t_1}\approx t^4/6$. Here the partition of length $t_1$ is free to translate backward from $\{a_j\}_{j=0}^{t_1-1}$ to $\{a_j\}_{j=-t_1+1}^{0}$ since $a_0$ should be in the $t_1$ partition, leading to $t_1$ degeneracy. With a given $t_1$, the number of configurations with three partitions of length $t_{2,3}$ is given by $\binom{t-t_1+2-1}{t-t_1}$. The overall factor of $t$ represents that the $\{a_j^*\}$ are free to translate over all $t$ basis states for $t$ degeneracy.

{\color{darkGreen} $\{2^21^{t-4}\}$.---}
Now we divide two copies of $t$ basis states $\{a_j\}$, $\{a_j^*\}$ into four partitions of length $t_j$ as $\{a_j\}=\{t_1t_2t_3t_4\}$ and $\{a_j^*\}=\{t_2t_1t_4t_3\}$ with a constraint $\sum_{j=1}^4t_j=t$~[Fig.~\ref{fig:Cycle}(c)]. The permutations $\sigma$ that connect partitions of the same length give rise to the $\{2^21^{t-4}\}$ cycle structure. To avoid overcounting, we define $t_1$ as the length of partition involving $a_0$.

The number of this type of permutations is given by $t\sum_{t_1\leq t}t_1\binom{t-t_1+3-1}{t-t_1}\approx t^5/24$. Here the partition of length $t_1$ is free to translate backward from $\{a_j\}_{j=0}^{t_1-1}$ to $\{a_j\}_{j=-t_1+1}^{0}$ since $a_0$ should be in the $t_1$ partition, leading to $t_1$ degeneracy. With a given $t_1$, the number of configurations with three partitions of length $t_{2,3,4}$ is given by $\binom{t-t_1+3-1}{t-t_1}$. The overall factor of $t$ represents that the $\{a_j^*\}$ are free to translate over all $t$ basis states for $t$ degeneracy.

Combining these two results, the $\sigma=\tau$ contribution to SFF is
\begin{align}
\langle K(t)\rangle\Big|_{\sigma=\tau}=\mathbb{E}[\zeta^2]^t\begin{aligned}[t]\bigg[t&+\frac{t^4}{6}\frac{\displaystyle \mathbb{E}[\zeta^6]}{\displaystyle \mathbb{E}[\zeta^2]^3}D^{-2}
\\
&+\frac{t^5}{24}\frac{\displaystyle \mathbb{E}[\zeta^4]^2}{\displaystyle \mathbb{E}[\zeta^2]^4}D^{-2}+\mathcal{O}(D^{-4})\bigg].\end{aligned}\label{eq:Diag}
\end{align}

\subsection{Off-diagonal contribution}
A more careful calculation involves the terms that exactly cancel the leading contribution of $\sigma=\tau$ in the unitary limit. Here we fix $\sigma$ as $\sigma \mathcal{T}^{-1}\sigma^{-1}\mathcal{T}$ forming cycle structures $=\{31^{t-3}\}$ or $\{2^21^{t-4}\}$, and list all possible $\tau\neq\sigma$ of the same order $\mathcal{O}(D^{-2})$ in the unitary case.

{\color{darkGreen} $\{31^{t-3}\}$.---}
One possible $\tau$ is the single elementary transposition acting on a pair of three basis states in the $3$-loop in $\{31^{t-3}\}$. There are three distinct choices, and the resulting $\sigma \mathcal{T}^{-1}\tau^{-1}\mathcal{T}$ forms cycle structure $\{21^{t-2}\}$. The other possible $\tau$ is two elementary transpositions acting on all three basis states of the $3$-loop in $\{31^{t-3}\}$, which has only one choice and leads to cycle structure $\sigma \mathcal{T}^{-1}\tau^{-1}\mathcal{T}=\{1^t\}$. The contributions of these two $\tau$ are
\begin{align}
\mathbb{E}[\zeta^2]^t\frac{t^4}{6}\bigg[3\frac{\text{Wg}(21^{t-2})}{\text{Wg}(1^t)}\frac{\text{Tr}[\zeta^4]}{\text{Tr}[\zeta^2]^2}+\frac{\text{Wg}(31^{t-3})}{\text{Wg}(1^t)}\bigg]\label{eq:OffDiagt4}
\\
=\mathbb{E}[\zeta^2]^t\frac{t^4}{6}\bigg[-3D^{-2}\frac{\mathbb{E}[\zeta^4]}{\mathbb{E}[\zeta^2]^2}+2D^{-2}\bigg].\nonumber
\end{align}

{\color{darkGreen} $\{2^21^{t-4}\}$.---}
One possible $\tau$ is the single elementary transposition acting on one of two $2$-loops in $\{2^21^{t-4}\}$. There are two distinct choices, and the resulting $\sigma \mathcal{T}^{-1}\tau^{-1}\mathcal{T}$ forms cycle structure $\{21^{t-2}\}$. The other possible $\tau$ is two elementary transpositions acting on two $2$-loops in $\{2^21^{t-4}\}$, which has only one choice and leads to cycle structure $\sigma \mathcal{T}^{-1}\tau^{-1}\mathcal{T}=\{1^t\}$. The contributions of these two $\tau$ are
\begin{align}
\mathbb{E}[\zeta^2]^t\frac{t^5}{24}\bigg[2\frac{\text{Wg}(21^{t-2})}{\text{Wg}(1^t)}\frac{\text{Tr}[\zeta^4]}{\text{Tr}[\zeta^2]^2}+\frac{\text{Wg}(2^21^{t-4})}{\text{Wg}(1^t)}\bigg]\label{eq:OffDiagt5}
\\
=\mathbb{E}[\zeta^2]^t\frac{t^5}{24}\bigg[-2D^{-2}\frac{\mathbb{E}[\zeta^4]}{\mathbb{E}[\zeta^2]^2}+D^{-2}\bigg].\nonumber
\end{align}

To sum up, from Eqs.~(\ref{eq:Diag}), (\ref{eq:OffDiagt4}), and (\ref{eq:OffDiagt5}), several leading contributions to SFF are given by Eq.~(\ref{eq:TotalSFF}).

\bibliography{ref}

\end{document}